\documentclass[10pt,twocolumn,twoside]{IEEEtran}
\usepackage{balance}

\usepackage[nocompress]{cite}
\usepackage{epsfig}
\usepackage{epstopdf}
\usepackage{graphicx}
\usepackage{float}
\usepackage{verbatim}
\usepackage[font=small]{caption}
\usepackage{hyperref}
\usepackage{wrapfig}
\usepackage[procnumbered,ruled,vlined]{algorithm2e}
\usepackage[noend]{algorithmic}
\usepackage{amsmath}
\usepackage{amssymb}
\usepackage{multirow}
\usepackage[table]{xcolor}
\usepackage{bm}
\usepackage{bbm}
\usepackage[labelformat=simple]{subcaption}

\DeclareMathOperator*{\argmin}{arg\,min}
\DeclareMathOperator*{\minimize}{minimize}
\newcommand{\diag}{\mathrm{diag}}
\newcommand{\HH}{\mathcal{H}}
\newcommand{\bHH}{{\bar{\mathcal{H}}}}
\newcommand{\GG}{\mathcal{G}}
\newcommand{\NN}{\mathcal{N}}
\newcommand{\LL}{\mathcal{L}}
\newcommand{\UU}{\mathcal{U}}
\newcommand{\FF}{\mathcal{F}}
\newcommand{\re}{\mathrm{Re}}
\newcommand{\im}{\mathrm{Im}}
\newcommand{\ii}{\mathbf{i}}
\newcommand{\YY}{\mathbf{Y}}
\newcommand{\MM}{\mathbf{M}}
\newcommand{\GY}{\mathbf{G}}
\newcommand{\BY}{\mathbf{B}}
\newcommand{\D}{\mathbf{D}}
\newcommand{\supp}{\mathrm{supp}}
\newcommand{\rank}{\mathrm{rank}}
\newcommand{\nul}{\mathrm{null}}
\newtheorem{obs}{Observation}
\newtheorem{lemma}{Lemma}
\newtheorem{corollary}{Corollary}

\begin{document}
\setlength{\textfloatsep}{2 pt}

\title{EXPOSE the Line Failures following\\ a Cyber-Physical Attack on the Power Grid}
\author{Saleh~Soltan,~\IEEEmembership{Member,~IEEE,}
        and~Gil~Zussman,~\IEEEmembership{Senior~Member,~IEEE}
\IEEEcompsocitemizethanks{\IEEEcompsocthanksitem S. Soltan is with the Department of Electrical Engineering, Princeton University, NJ, 08544, and G. Zussman is with the Department
of Electrical Engineering, Columbia University, New York,
NY, 10027.\protect\\
E-mails: ssoltan@princeton.edu, gil@ee.columbia.edu}}

\maketitle
\begin{abstract}
Recent attacks on power grids demonstrated the vulnerability of the grids to cyber and physical attacks. To analyze this vulnerability, we study cyber-physical attacks that affect both the power grid physical infrastructure and its underlying Supervisory Control And Data Acquisition (SCADA) system. We assume that an adversary attacks an area by: (i) disconnecting some lines within that area, and (ii) obstructing the information (e.g., status of the lines and voltage measurements) from within the area to reach the control center. We leverage the algebraic properties of the AC power flows to introduce the efficient EXPOSE Algorithm for detecting line failures and recovering voltages inside that attacked area after such an attack. The EXPOSE Algorithm outperforms the state-of-the-art algorithm  for detecting line failures using partial information under \emph{the AC power flow model} in terms of scalability and accuracy. The main advantages of the EXPOSE Algorithm are that its running time is independent of the size of the grid and number of line failures, and that it provides accurate information recovery under some conditions on the attacked area. Moreover, it approximately recovers the information and provides the confidence of the solution when these conditions do not hold.
\end{abstract}


\section{Introduction}\label{sec:Introduction}

Recent cyber attack on the Ukrainian grid in December 2015~\cite{UkraineBlackout} demonstrated  the vulnerability of  power grids to cyber attacks. As indicated in the aftermath report of the attack~\cite{UkraineBlackout}, once the attackers obtain access to the grid's Supervisory Control And Data Acquisition (SCADA) system, they can delete, modify, and spoof the data as well as remotely change the grid's topology by activating the circuit breakers.

The power grid infrastructure is also vulnerable to physical attacks. Such an attack occured in April 2014 in San Jose, California, when snipers tried to shut down a substation simply by shooting at its transformers~\cite{sniper2014}. Hence, a physical attack on the power lines and the measurement devices can have a similar affect to a cyber attack.

\begin{figure}[t]
\centering
\includegraphics[scale=0.65]{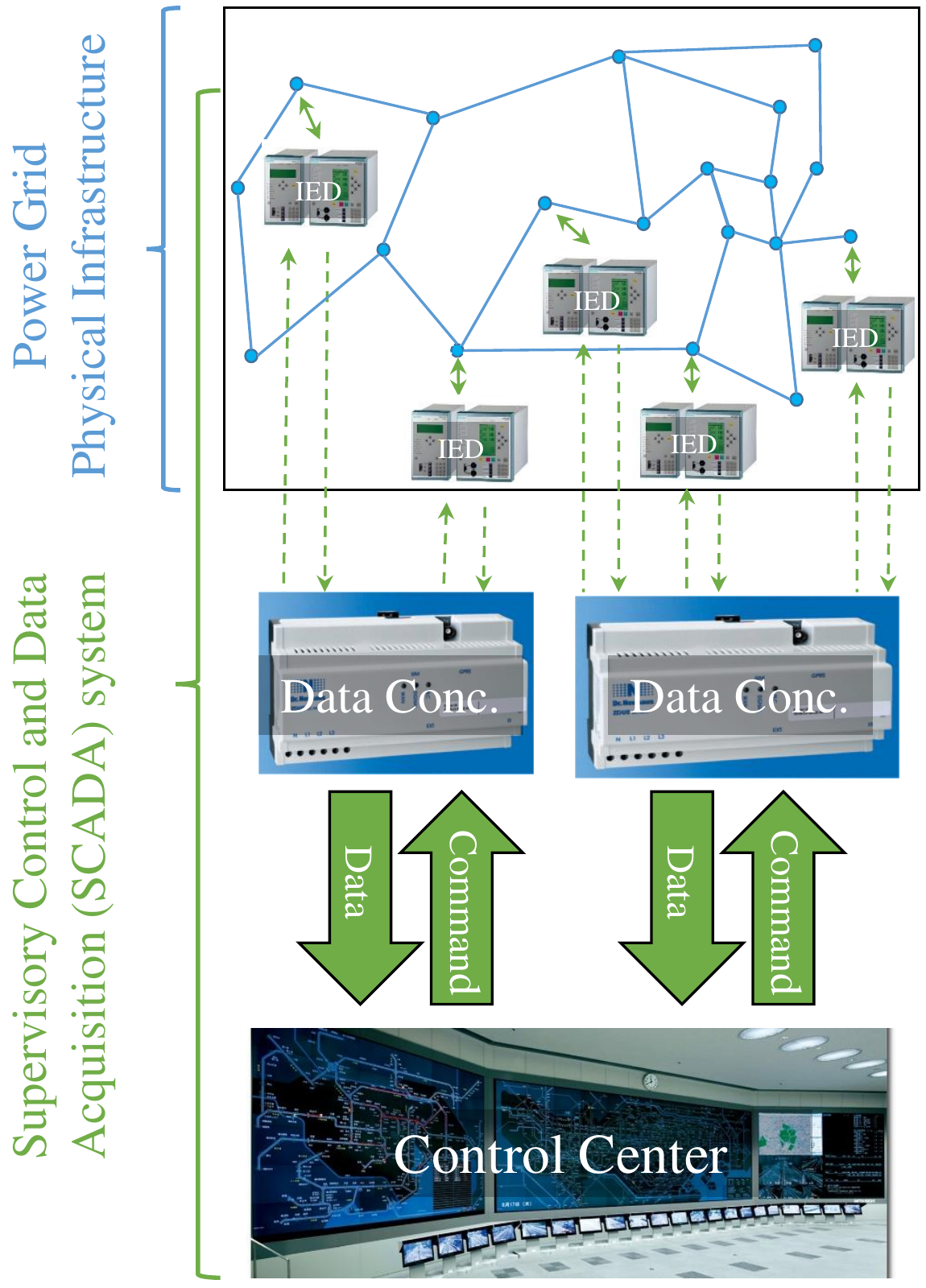}
\caption{The main components of the power grid. The Intelligent Electronic Devices (IED) measure voltage magnitudes and phase angles, and send these information via Data Concentrators to the Control Center.} 
\label{fig:components}
\end{figure}

To analyze these vulnerabilities,  in this paper, we study cyber-physical attacks that affect both the power grid physical infrastructure and its SCADA system. Fig.~\ref{fig:components} shows the main components of the power grids. An adversary can attack the grid by damaging the power lines and measurement devices with a physical attack, by remotely disconnecting the lines and erasing the measurements data with a cyber attack, or by performing a combination of the both.

\begin{figure*}[t]
\centering
\includegraphics[scale=0.45]{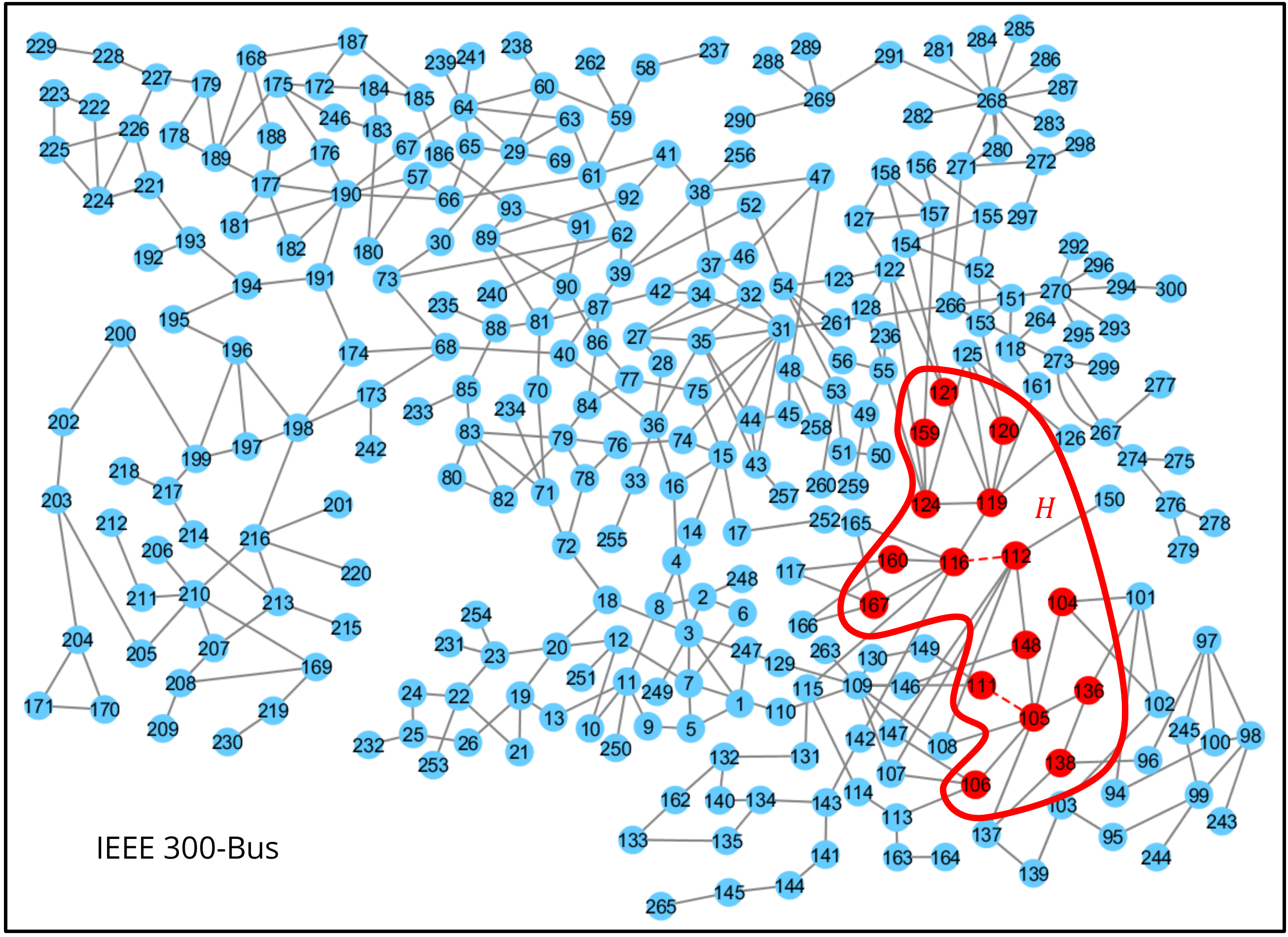}
\caption{The attack model. An adversary attacks a zone by disconnecting some of its lines (red dashed lines) and disallowing the information from within the zone to reach the control center. $\GG$ is the power grid graph and $\HH$ is a subgraph of $\GG$ that represents the  attacked zone.} 
\label{fig:attack_300MA}
\end{figure*}

Independent of the attack strategy, we assume that an adversary attacks an area by: (i) disconnecting some lines within that area (\emph{failed lines}), and (ii) obstructing the information (e.g., status of the lines and voltage measurements) from within the area to reach the control center. We call this area, the \emph{attacked zone}. Our objective is to detect the failed lines and recover the voltages inside the attacked zone using the information available outside of the attacked zone as well as the information before the attack. An example of such an attack on the IEEE 300-bus system is depicted in Fig.~\ref{fig:attack_300MA}.

 We studied a similar attack scenario in~\cite{soltan2016state} using the linearized DC power flows. In a recent extension~\cite{soltan2017power}, the methods in~\cite{soltan2016state} were modified to statistically recover the information under the AC power flows. However, due to the inaccuracy of the DC power flows, the methods in~\cite{soltan2017power} could not guarantee the correct information recovery under the AC power flows.

In this paper, we \emph{directly leverage the properties of the nonlinear AC power flows} to detect the line failures and recover the voltages after an attack with guarantee of performance. In particular, we prove that if there is a matching between the nodes (buses) inside and outside of the attacked zone that covers all the nodes inside the attacked zone, the voltages can be accurately recovered by solving a set of linear equations. Moreover, given the successful recovery of the voltages, we prove that if the attacked zone is acyclic (i.e., lines in the attacked zone do not form any cycles), then the failed lines can be accurately detected by solving a set of linear equations.

We extend these results and show that given the successful recovery of the voltages, the failed lines can still be accurately detected  by solving a Linear Program (LP), even if the attacked zone is not acyclic. We further show that even if there is no matching between the nodes inside and outside of the attacked zone that covers all the inside nodes and the attacked zone is not acyclic, one can still approximately recover the voltages and detect the line failures using convex optimization.

Based on the results, we then introduce the EXPress line failure detection using partially ObSErved information (EXPOSE) Algorithm. It outperforms the state-of-the-art algorithm  for detecting line failures using partial information under the AC power flows in terms of scalability and accuracy. The main advantages of the EXPOSE Algorithm are that its running time is independent of the size of the grid and number of line failures, and that it provides accurate information recovery under some conditions on the attacked zone. Moreover, it approximately recovers the information and provides the confidence of the solution when these conditions do not hold.

Most of the related work rely on the DC power flows and deploy brute force search approaches. These approaches do not scale well, and therefore, are limited only to detecting single and double line failures using partial measurements~\cite{tate2008line,tate2009double,zhu2012sparse,zhao2012pmu,zhu2014phasor}. To represent these approaches and for comparison purposes only, we also introduce a naive Brute Force Search (BFS) Algorithm for detecting line failures after the attack.

Finally, while we analytically prove that the EXPOSE Algorithm guarantees to accurately recover the voltages and detect line failures under some conditions, we also numerically evaluate  its performance when those conditions do not hold. In particular, we evaluate the performance of the EXPOSE Algorithm as the attacked zone becomes topologically more complex and compare its running time to the BFS Algorithm by considering all single, double, and triple line failures in 5 nested attacked zones. Based on the simulation results, we conclude that despite its
accuracy, the BFS Algorithm is not practical for line failures
detection in large networks and that the EXPOSE Algorithm can provide relatively accurate results
exponentially faster. For example, the EXPOSE algorithm recovers the voltages with less than 15\% error and detects line failures with less than 1 false negative on average, after all single, double, and triple line failures in an attacked zone that satisfies none of the conditions for the accuracy of the EXPOSE Algorithm.

\section{Related Work}\label{sec:Related}
Vulnerability of power grids to failures and attacks has been widely studied~\cite{ashok2017cyber,pinar_power,kim2016analyzing,liu2014distributed,Bern2012ACM,Dobson,bienstock2016electrical,soltan2016analyz,srivastava2013modeling,pasqualetti2011cyber}. In particular false data injection attacks on power grids and anomaly detection have been studied using the DC power flows in~\cite{kim2013topology,liu2011false,dan2010stealth,vukovic2011network,li2015quickest,kim2015subspace}. These studies focused on the observability of the failures and attacks in the grid.

The problem studied in this paper is similar to the problem of line failures detection using phase angle measurements~\cite{tate2008line,tate2009double,garcia2016line,zhu2012sparse}. Up to two line failures detection, under the DC power flow model, was studied in~\cite{tate2008line,tate2009double}. Since the provided methods in~\cite{tate2008line,tate2009double} are greedy-based methods that need to search the entire failure space, the running time of these methods grows exponentially as the number of failures increases. Hence, these methods cannot be generalized to detect higher order failures.  Similar greedy approaches with likelihood detection functions  were studied in~\cite{manousakis2012taxonomy,Khandeparkar2014Eff,zhao2012pmu,zhao2014identification,zhu2014phasor} to address the PMU placement problem under the DC power flow model. 

 The problem of line failures detection in an internal system using the information from an external system was also studied in~\cite{zhu2012sparse} based (again) on the DC power flow model. The proposed algorithm works for only one and two line failures, since it depends on the sparsity of line failures. 


 In a recent work~\cite{garcia2016line}, a linear multinomial regression model was proposed as a classifier for a single line failure detection using transient voltage phase angles data. Due to the time complexity of the learning process for multiple line failures, this method is impractical for detecting higher order failures. Moreover, the results provided in~\cite{garcia2016line} are empirical with no performance guarantees.

Finally, in a recent series of works, the vulnerability of power grids to undetectable cyber-physical attacks is studied~\cite{li2016bilevel,deng2017ccpa,zhang2016physical} using the DC power flows. These studies are mainly focused on designing attacks that affect the entire grid and therefore may remain undetected.

To the best of our knowledge, our methods presented in this paper and \cite{soltan2017power} are the only methods for line failures detection under the AC power flows that can be used to detect any number of line failures and scale well with size of the grid. However, the EXPOSE Algorithm provided in this paper is more accurate than the method provided in~\cite{soltan2017power}.

\section{Model and Definitions}\label{sec:model}
\subsection{AC Power Flow Equations}
 A power grid with {$n$}  nodes (buses) and {$m$}    transmission lines    can be represented by an undirected graph $\mathcal{G}(\mathcal{N},\mathcal{L})$, where $\mathcal{N}=\{1,2,\dots,n\}$ denotes the set of nodes  and $\mathcal{L}=\{l_1,l_2,\dots,l_m\}$ denotes the set of lines or edges. In the steady-state, the status of each node $i$ is represented by its voltage $V_i=|V_i| e^{\mathbf{i}\theta_i}$ in which $|V_i|$ is the voltage magnitude, $\theta_i$ is the voltage phase angle, and $\mathbf{i}$ denotes the imaginary unit.

 The goal of the AC power flow analysis is the computation of the voltage magnitudes and phase angles at each bus in   steady-state conditions~\cite{glover2012power}. In the steady-state, the AC power flow equations can be written in matrix form as follows:
\begin{align}
&\mathbf{Y}V=I,\label{eq:YVI}\\
&S=\diag(V)I^*,\label{eq:SVI}
\end{align}
where $^*$ denotes the complex conjugation, $ V=[V_1, \dots, V_n]^{\mathrm{T}}$ is the vector of node voltages, $I=[I_1,I_2,\dots,I_n]^T$ is the vector of injected node currents, $S=[S_1,S_2,\dots,S_n]^T$ is the vector of injected apparent powers, and $\YY$ is the admittance matrix of the graph.

The elements of the admittance matrix $\mathbf Y$ which depends on the topology of the grid as well as the admittance values of the lines, is defined as follows:
\begin{equation*} \label{eq:Y_bus}
  Y_{ik}= \begin{cases}
  y_{ii}+\sum_{i \neq k} {y_{ik}}, & \mbox{if } k=i \\
  -y_{ik},  & \mbox{if } k\in N(i)\\
  0, & \mbox{if}~ k\notin N(i)
\end{cases}
\end{equation*}
where $N(i)$ denotes the direct neighbors  of node $i$, $y_{ik} $ is the  equivalent  admittance of the lines  from node $i$ to $k$, and $y_{ii}$ is sum of the \emph{shunt admittances} at node $i$. In this paper, we assume that the shunt admittances are negligible, and therefore, $y_{ii}=0$ for all $i\in \mathcal{N}$. The admittance matrix can also be written in term of its real and imaginary parts as $\mathbf{Y}=\mathbf{G}+\mathbf{i} \mathbf{B}$ where $\mathbf{G}$ and $\mathbf{B}$ are real matrices. Using this and the definition of the apparent power $S_{i}=P_{i} +\mathbf{i}  Q_{i}$ in~(\ref{eq:YVI}-\ref{eq:SVI}) results in the equations for the active power $P_i$ and the reactive power $Q_i$ at each node $i$ as well.

\subsection{Incidence Matrix}\label{subsec:incidence}

Under an arbitrary direction assignment to the edges of $\GG$, the \emph{incidence matrix} of $\GG$ is denoted by $\D\in\{-1,0,1\}^{n\times m}$ and defined as,
\begin{equation*}
d_{ij}=
\begin{cases}
0&\text{if}~l_j~\text{is not incident to node}~i,\\
1&\text{if}~l_j~\text{is coming out of node}~i,\\
-1&\text{if}~l_j~\text{is going into node}~i.
\end{cases}
\end{equation*}
For each line $l_j=(i,k)$, define $y_{l_j}:=y_{ik}$. It can be verified that $\YY=\D\diag([y_{l_1},y_{l_2},\dots, y_{l_m}])\D^T$. As we demonstrate in Section~\ref{sec:state_estimate}, the incidence matrix is a very useful matrix for detecting line failures in power grids.
\subsection{Basic Graph Theoretical Terms}\label{subsec:graph}

\noindent \textbf{Matching:} A \emph{matching} in a graph is a set of pairwise nonadjacent edges. If $\mathcal{M}$ is a matching,
the two ends of each edge of $\mathcal{M}$ are said to be \emph{matched} under $\mathcal{M}$, and each vertex
incident with an edge of $\mathcal{M}$ is said to be \emph{covered} by $\mathcal{M}$.

\noindent \textbf{Cycle:} A \emph{cycle} in a graph is a sequence of its distinct nodes $u_1,u_2,\dots,u_k$ such that for all $i<k$, $\{u_i,u_{i+1}\}\in \mathcal{L}$, and also $\{u_k,u_1\}\in \mathcal{L}$. A graph with no cycle is called \emph{acyclic}.

\subsection{Attack Model}\label{subsec:attack}
We assume that an adversary attacks an area by: (i) disconnecting some lines within that area (\emph{failed lines}), and (ii) obstructing all the information (e.g., status of the lines and voltage measurements) from within the area to reach the control center. We call this area, the \emph{attacked zone}. 

Fig.~\ref{fig:attack_300MA} shows an example of an attack on the area represented by $\HH=(\NN_\HH,\LL_\HH)$. 
We denote the set of failed lines in the attacked zone $\HH$ by $\FF\subseteq \LL_\HH$. Upon failure, the failed lines are removed from the graph and the flows are redistributed according to the AC power flows.
\emph{Our objective is to estimate the voltages and detect the failed lines inside the attacked zone using the changes in the voltages outside of the zone.}

We use the prime symbol $(')$ to denote the values after an attack (e.g, $\YY'$ denotes the admittance matrix of the grid and $V'$ denotes node voltages after the attack).
Using this notation, if $\bar{\HH}$ denotes the set of nodes outside of the attacked zone, then given $V_{\bar{\HH}}'$ and $S'$, we want to recover $V_{\HH}'$ and $\FF$. Notice that it is reasonable to assume that we know $S'_\HH$ after the attack, since for the load nodes inside that attacked zone $\HH$, we can assume that $P'$ and $Q'$ remain similar to their values before the attack and for the generators we can assume that generator operators can safely report the generated $P'$ and $Q'$ values. Equivalently, we can assume that $P$ and $Q$ do not change much after the attack and therefore $S'=S$.

Detecting line failures after such an attack is crucial in maintaining the stability of the grid, since it may result in further line overloads and failures, if the proper load shedding mechanism is not applied. An effective load shedding requires the exact knowledge of the topology of the grid.



%
\textbf{Notation.} For any complex number $x = \re(x)+\ii~\im(x)$, real numbers $\re(x)$ and $\im(x)$ denote its real and imaginary values, respectively. For a vector $X$, $\supp(X)$ denotes the set of its nonzero entries. If $\HH_1,\HH_2$ are two subgraphs of $\GG$, $\YY_{\HH_1|\HH_2}$ denotes the submatrix of $\YY$ with rows from $V_{\HH_1}$ and columns from $V_{\HH_2}$. Moreover, $\YY_{\HH_1}$ denotes the submatrix of $\YY$ with all the rows associated with $V_{\HH_1}$. For instance, $\YY$ can be written in any of the following forms,
\begin{equation*}
\YY=
\begin{bmatrix}
\YY_{\HH|\HH}&\YY_{\HH|\bHH}\\
\YY_{\bHH|\HH}&\YY_{\bHH|\bHH}
\end{bmatrix}
, \YY=
\begin{bmatrix}
\YY_{\HH}\\
\YY_{\bHH}
\end{bmatrix}.
\end{equation*}

\section{State Estimation}\label{sec:state_estimate}
In this section, we provide the analytical building blocks of the EXPOSE Algorithm which can be used to estimate the state of the grid following a cyber-physical attack. Notice that the state estimation problem considered here is different from the classical state estimation problem in power grids. Here, besides estimating the voltage magnitudes and phase angles in the attacked area, the algorithm needs to estimate the topology of the grid as well.
\subsection{Voltage Recovery}\label{subsec:voltage_recover}
Here, we provide a method to recover the voltages inside that attacked zone after the attack.
\begin{obs}\label{lem:admittance_matrix}
The admittance matrix of the grid does not change outside of the attacked zone (i.e., $\YY_{\bHH}=\YY_{\bHH}'$).
\end{obs}
\begin{IEEEproof}
Since the line failures only happen inside $\HH$, following the definition of the admittance matrix (see Section~\ref{sec:model}), after the attack only the entries of $\YY_{\HH|\HH}$ change. Hence, $\YY_{\bHH}$ remains unchanged.
\end{IEEEproof}

From Observation~\ref{lem:admittance_matrix} and using (\ref{eq:YVI}), we have:
\begin{align}
&\YY_{\bHH}'V'=I_\bHH'\Rightarrow \YY_{\bHH}V'=I_\bHH'\Rightarrow \YY_{\bHH}^*V'^*=I_\bHH'^*\nonumber\\
\Rightarrow& \diag(V'_\bHH)\YY_{\bHH}^*V'^*=\diag(V'_\bHH)I_\bHH'^*\nonumber\\
\Rightarrow& \diag(V'_\bHH)\YY_{\bHH}^*V'^*=S_\bHH'\nonumber\\
\Rightarrow& \diag(V'_\bHH)\YY_{\bHH|\bHH}^*V_\bHH'^*+\diag(V'_\bHH)\YY_{\bHH|\HH}^*V_\HH'^*=S_\bHH'.\label{eq:v_recovery_S}
\end{align}
Notice that in (\ref{eq:v_recovery_S}) all the variables are known after the attack except $V_\HH'^*$. Define $E_\bHH := -\YY_{\bHH|\bHH}^*V_\bHH'^*+\diag(V_\bHH'^{-1})S_\bHH'$ which can be computed from the given variables after the attack. Then, we can separate the real and imaginary parts of (\ref{eq:v_recovery_S}) using block matrices as follows:
\begin{align}
\begin{bmatrix}
\GY_{\bHH|\HH}&-\BY_{\bHH|\HH}\\
\BY_{\bHH|\HH}&\GY_{\bHH|\HH}
\end{bmatrix}
\begin{bmatrix}
\re(V_\HH')\\
\im(V_\HH')
\end{bmatrix}
&=
\begin{bmatrix}
\re(E_\bHH)\\
-\im(E_\bHH)
\end{bmatrix}.\label{eq:v_recovery_matrix}
\end{align}
One can see that $\re(V_\HH')$ and $\im(V_\HH')$ can be uniquely recovered if the matrix on the left hand side of (\ref{eq:v_recovery_matrix}) has full column rank. The following lemma provides the connection between the rank of that matrix and the topology of the grid.
\begin{lemma}\label{lem:matching}
If there is matching between the nodes in $\bHH$ and $\HH$ that covers the nodes in $\HH$, then the following matrix has full column rank almost surely,
\begin{equation*}
\MM:=\begin{bmatrix}
\GY_{\bHH|\HH}&-\BY_{\bHH|\HH}\\
\BY_{\bHH|\HH}&\GY_{\bHH|\HH}
\end{bmatrix}.
\end{equation*}
\end{lemma}
\begin{IEEEproof}
Suppose $\UU\subseteq \NN_{\bHH}$ are the matched nodes which are in $\bHH$. Since the matching covers $\HH$, thus $|\UU|=|\NN_{\HH}|$. To show that $\MM$ has full column rank, we show that
\begin{equation*}
\det(\MM_{U|H}):=\det\begin{bmatrix}
\GY_{\UU|\HH}&-\BY_{\UU|\HH}\\
\BY_{\UU|\HH}&\GY_{\UU|\HH}
\end{bmatrix} \neq 0,
\end{equation*}
almost surely. $\det(\MM_{\UU|\HH})$ can be considered as a polynomial in terms of the entries of $\MM_{\UU|\HH}$ using Leibniz formula. Now assume $\UU=\{u_1,u_2,\dots,u_{|\NN_{\HH}|}\}$ are matched to $\NN_{\HH}=\{v_1,v_2,\dots,v_{|\NN_{\HH}|}\}$ in order. It can be seen that $\prod_{i=1}^{|\NN_{\HH}|} G_{u_iv_i}^2$ and $\prod_{i=1}^{|\NN_{\HH}|} B_{u_iv_i}^2$ are two terms with nonzero coefficient in $\det(\MM_{\UU|\HH})$. Therefore, $\det(\MM_{\UU|\HH})$ is a nonzero polynomial in terms of its entries. Now since the set of roots of a nonzero polynomial is a measure zero set in the real space, thus $\det(\MM_{\UU|\HH})\neq0$ almost surely.
\end{IEEEproof}
\begin{corollary}
If there is matching between the nodes in $\bHH$ and $\HH$ that covers the nodes in $\HH$, then $V'_\HH$ can be recovered almost surely.
\end{corollary}
\subsection{Line Failures Detection}\label{subsec:detect_lines}
Assume $V'_\HH$ is successfully recovered using (\ref{eq:v_recovery_matrix}). In this subsection, using $V'_\HH$, we provide a method to detect the set of line failures $\FF$.
\begin{lemma}\label{lem:DX}
There exists a complex vector $X\!\in\!\mathbb{C}^{|\LL_\HH|}$ such that
\begin{equation}\label{eq:DX}
\YY_\HH V'=I_\HH'+\D_{\HH} X,
\end{equation}
and $\supp(X)=\FF$. Moreover, the vector $X$ is unique if, and only if, $\D_{\HH}$ has full column rank.
\end{lemma}
\begin{IEEEproof}
Without loss of generality assume $\FF=\{l_1,l_2,\dots,l_k\}$. It can be seen that $\YY'=\YY-\D\diag([y_{l_1},y_{l_2},\dots, y_{l_k},0,0,\dots,0])\D^T$. Hence,
\begin{align*}
\YY'V'&=\YY V'-\D\diag([y_{l_1},y_{l_2},\dots, y_{l_k},0,0,\dots,0])\D^T V'\\
I' &= \YY V'-\D\diag([y_{l_1},y_{l_2},\dots, y_{l_k},0,0,\dots,0])\D^T V'.
\end{align*}
Now if we only focus on the rows associated with the nodes in $\HH$, it can been seen that
\begin{align*}
I_{\HH}' &= \YY_{\HH} V'-\D_{\HH}\diag([y_{l_1},y_{l_2},\dots, y_{l_k},0,0,\dots,0])\D_{\HH}^T V'.
\end{align*}
Hence, vector $X:=\diag([y_{l_1},y_{l_2},\dots, y_{l_k},0,0,\dots,0])\D_{\HH}^T V'$ satisfies (\ref{eq:DX}). It can also be seen that only the entries of $X$ that are associated with the failed lines are nonzero and therefore $\supp(X)=\FF$. In order for (\ref{eq:DX}) to have a unique solution, $\D_{\HH}$ should have full column rank.
\end{IEEEproof}
\begin{corollary}\label{cor:fail_detect}
There exist a real vector $X\in\mathbb{R}^{|\LL_\HH|}$ such that
\begin{equation}\label{eq:fail_detect}
\re\{\YY_\HH^* V'^*\} = \re\{\diag(V_\HH')^{-1}S_\HH'\}+\D_{\HH} X,
\end{equation}
and $\supp(X)=\FF$. Moreover, the vector $X$ is unique if, and only if, $\D_{\HH}$ has full column rank.
\end{corollary}
\begin{IEEEproof}
Using (\ref{eq:SVI}) and Lemma~\ref{lem:DX} gives the result.
\end{IEEEproof}
Corollary~\ref{cor:fail_detect} indicates that the set of line failures can be detected by solving a matrix equation, if $\D_{\HH}$ has full column rank. The following lemma provides the connection between the rank of that matrix and the topology of the attacked zone.
\begin{lemma}
The solution vector $X$ to (\ref{eq:fail_detect}) is unique if, and only if, $\HH$ is acyclic.
\end{lemma}
\begin{IEEEproof}
It is easy to verify that (\ref{eq:fail_detect}) has a unique solution $X$ if, and only if, $\D_{\HH}$ has linearly independent columns. On the other hand, it is known in graph theory that  $\rank(\D_{\HH})=|\NN_{\HH}|-c$ in which $c$ is the number of connected components of $\HH$~\cite[Theorem 2.3]{bapat2010graphs}. Therefore, $\D_{\HH}$ has linearly independent columns if and only if $\LL_{\HH}=|\NN_{\HH}|-c$ which means that each connected component of $\D_{\HH}$ is acyclic.
\end{IEEEproof}
\begin{corollary}\label{cor:asyclic}
If $\HH$ is acyclic, then the set of line failures $\FF$ can be detected by solving (\ref{eq:fail_detect}) for $X$.
\end{corollary}
Corollary~\ref{cor:asyclic} states that the set of line failure can accurately be detected if $\HH$ is acyclic. The importance of this result is in demonstrating that the set of line failures can be efficiently detected by solving a matrix equation, independent of the number of line failures.

We can use a similar idea as in~\cite{soltan2016state} to extend this approach to when $\HH$ is not acyclic. If we assume that the set of line failures are sparse compare to the total number of lines in $\HH$, we can detect line failures by finding the solution of following optimization problem instead:
\begin{align}
&\minimize_{X\in\mathbb{R}^{|\LL_\HH|}} \|X\|_1~\text{s.t.}\nonumber\\
&\re\{\YY_\HH^* V'^*\} = \re\{\diag(V_\HH')^{-1}S_\HH'\}+\D_{\HH} X. \label{eq:LP_fail_detect}
\end{align}
Notice that optimization problem (\ref{eq:LP_fail_detect}) can be solved efficiently using Linear Programming (LP).
\begin{lemma}\label{lem:cycle}
If $\HH$ is a cycle and less than half of its edges are failed, then the solution $X$ to the optimization problem (\ref{eq:LP_fail_detect}) is unique and $\supp(X)=\FF$.
\end{lemma}
\begin{IEEEproof}
The idea of the proof is similar to the idea used in the proof in~\cite[Lemma 4]{soltan2016state}. Here without loss of generality, we assume that $\D_{\HH}$ is the incidence matrix of $\HH$ when edges of the cycle are directed clockwise. Since $\HH$ is connected, it is known that $\rank(\D_{\HH})=|\NN_{\HH}|-1$~\cite[Theorem 2.2]{bapat2010graphs}. Therefore, $\dim(\nul(\D_{\HH}))=1$. Suppose $\mathbbm{1}\in \mathbb{R}^{|\LL_{\HH}|}$ is the all one vector. It can be seen that $\D_{\HH}\mathbbm{1}=0$. Since $\dim(\nul(\D_{\HH}))=1$, $\mathbbm{1}$ is the basis for the null space of $\D$. Suppose $X$ is a solution to (\ref{eq:LP_fail_detect}) such that $\supp(X)=\FF$ (from Lemma~\ref{lem:DX} we know that such a solution exists). To prove that $X$ is the unique solution for (\ref{eq:LP_fail_detect}), we prove that $\forall c\in\mathbb{R}\backslash\{0\}$, $\|X\|_1<\|X-c\mathbbm{1}\|_1$. Without loss of generality we can assume that $x_1,x_2,\dots,x_k$ are the nonzero elements of $X$, in which $k=|\FF|$. From the assumption, we know that $k<|\LL_{\HH}|/2$. Hence,
\begin{align*}
\|X-c\mathbbm{1}\|_1 &= \sum_{i=1}^k|x_i-c|+(|\LL_{\HH}|-k)|c|\\
&=\sum_{i=1}^k(|x_i-c|+|c|)+(|\LL_{\HH}|-2k)|c|\\
&\geq\sum_{i=1}^k|x_i|+(|\LL_{\HH}|-2k)|c|>\sum_{i=1}^k|x_i|=\|X\|_1.
\end{align*}
Thus, the solution to (\ref{eq:LP_fail_detect}) is unique and $\supp(X)=\FF$.
\end{IEEEproof}

Lemma~\ref{lem:cycle} can be extended to planar graphs similar to the result for the DC power flows presented in~\cite[Theorem 2]{soltan2016state}. The proof and the argument in~\cite{soltan2016state} should apply with a very slight change here as well. To avoid repetition, we do not present a similar Lemma here.
\subsection{Simultaneous Recovery and Detection}\label{subsec:simul}
In order to extend our approach to the cases that (\ref{eq:v_recovery_matrix}) does not have a unique solution, we can solve (\ref{eq:v_recovery_matrix}) and (\ref{eq:LP_fail_detect}) at the same time. Therefore, in order to recover the voltages and detect the line failures at the same time, one needs to solve the following optimization problem:
\begin{align}
&\minimize_{X\in\mathbb{R}^{|\LL_\HH|},V_\HH'\in\mathbb{C}^{|\NN_\HH|}} \|X\|_1~\text{s.t.}\nonumber\\
&\GY_{\bHH|\HH}\re(V_\HH')-\BY_{\bHH|\HH}\im(V_\HH') = \re(E_\bHH) \nonumber\\
&\BY_{\bHH|\HH}\re(V_\HH')+\GY_{\bHH|\HH}\im(V_\HH') = -\im(E_\bHH)\nonumber\\
&\re\{\YY_\HH^* V'^*\} = \re\{\diag(V_\HH')^{-1}S_\HH'\}+\D_{\HH} X. \label{eq:Quad_Simul}
\end{align}

However, since $V_\HH'$ is part of the variables, this optimization problem is not linear and convex anymore. To resolve this issue, we need to approximate $\diag(V_\HH')^{-1}$ with a linear function in terms of $V_\HH'$. For this, we have:
\begin{align*}
\diag(V_\HH')^{-1}\!=\!\diag(|V_\HH'|)^{-2}\!\big(\diag(\re(V_\HH'))\!-\!\mathbf{i}~\diag(\im(V_\HH'))\big).
\end{align*}
On the other hand, the voltage magnitudes are almost constant at each node before and after the failure ($|V_\HH'|\approx |V_\HH|$), hence:
 \begin{align*}
\diag(V_\HH')^{-1}\!\approx\!\diag(|V_\HH|)^{-2}\!\big(\diag(\re(V_\HH'))\!-\!\mathbf{i}~\diag(\im(V_\HH'))\big).
\end{align*}
We can use the approximation above in optimization (\ref{eq:Quad_Simul}) in order to relax its nonconvexity. 
Notice that since optimization (\ref{eq:Quad_Simul}) is for the cases in which the solution to (\ref{eq:v_recovery_matrix}) is not unique and therefore the voltages cannot be recovered uniquely, some conditions should be placed on the voltages such that the recovered voltages are near operable conditions. To do so, we add a convex constraint on the voltage magnitudes of the nodes in $\HH$ after the attack as $|V_\HH'|\leq 1.1\mathbbm{1}_\HH$, in which $\mathbbm{1}_\HH$ is an all ones vector of size $|\NN_\HH|$. Hence, the following convex optimization can be used to detect the set of line failures and recover the voltages when the solution to (\ref{eq:v_recovery_matrix}) is not unique:
\begin{align}
&\minimize_{X\in\mathbb{R}^{|\LL_\HH|},V_\HH'\in\mathbb{C}^{|\NN_\HH|}} \|X\|_1~\text{s.t.}\nonumber\\
&\GY_{\bHH|\HH}\re(V_\HH')-\BY_{\bHH|\HH}\im(V_\HH') = \re(E_\bHH) \nonumber\\
&\BY_{\bHH|\HH}\re(V_\HH')+\GY_{\bHH|\HH}\im(V_\HH') = -\im(E_\bHH)\nonumber\\
&|V_\HH'|\leq 1.1\mathbbm{1}_\HH\nonumber\\
&\re\{\YY_\HH^* V'^*\} \approx \diag(|V_\HH|)^{-2}\diag(\re(V_\HH'))P'_\HH \nonumber \\
&~~~~~+\diag(|V_\HH|)^{-2}\diag(\im(V_\HH'))Q'_\HH+\D_{\HH} X. \label{eq:conv_Simul_v}
\end{align}
Notice that to enforce $AA\approx BB$ for arbitrary vectors $AA$ and $BB$, we use convex constraint $\|AA-BB\|_2<\epsilon$ for a small value $\epsilon$. In Section~\ref{sec:NumRes}, we evaluate the accuracy of the results obtained by solving the convex optimization problem (\ref{eq:conv_Simul_v}) as part of the EXPOSE Algorithm.

\begin{algorithm}[t]
\caption{EXPress line failure detection using partially ObSErved information (EXPOSE)}
\label{algorithm:EXPOSE}
\small
\begin{trivlist}
\item\textbf{Input:} A connected graph $\GG$, attacked zone $\HH$, $V$, $V'_{\bar{\HH}}$, and $S'$ 
\end{trivlist}
\begin{algorithmic}[1]
\IF {$\MM$ is full rank}
    \STATE Solve (\ref{eq:v_recovery_matrix}) to recover $V_{\HH}'$
    \STATE Use the recovered $V_{\HH}'$ in (\ref{eq:LP_fail_detect}) to find the solution vector $X$
\ELSE
    \STATE Find the solution $V_{\HH}'$  and $X$ to the optimization (\ref{eq:conv_Simul_v})
\ENDIF
\STATE Compute $\FF = \supp(X)$
\STATE Compute the confidence of the solution $c_P$ and $c_Q$
\RETURN $V_{\HH}',\FF,c_P,$ and $c_Q$
\end{algorithmic}
\end{algorithm}

\subsection{Confidence of the Solution}\label{subsec:conf_sol}

Once the set of line failures is detected and the voltages are recovered, one can compute the confidence of the solution using (\ref{eq:YVI}-\ref{eq:SVI}). Assume $\YY^\dagger$ and $V^\dagger$ denote the admittance matrix of the grid after removing the detected lines and the recovered voltages after the attack, respectively. If the detection and recovery are done correctly, then $\re\{\diag(V^\dagger)^* \YY V^\dagger\} = P'$ and $\im\{\diag(V^\dagger)^* \YY V^\dagger\} = Q'$. However, if the detection and recovery are not done correctly, this equalities do not hold. We can use the difference between the two sides of these equalities as a measure for the correctness of the solution.

We define $c_P$ and $c_Q$ to denote the confidence of the solution based on $P$ and $Q$ as follows:
\begin{align}
\!\!c_P\! :=\! (1\!-\!\|\re\{\diag(V^\dagger)^* \YY V^\dagger\}\! -\! P'\|_2/\|P'\|_2)^+\!*\!100\label{eq:confidence_P},\\
\!\!c_Q \!:=\! (1\!-\!\|\im\{\diag(V^\dagger)^* \YY V^\dagger\}\! -\! Q'\|_2/\|Q'\|_2)^+\!*\!100\label{eq:confidence_Q},
\end{align}
in which $(x)^+ := \max(0,x)$. If $c_P,c_Q\approx 100\%$, then it means that the solution is reliable. If not, depending on the $c_P$ or $c_Q$ values, one can see how reliable the solution actually is.

\section{EXPOSE Algorithm}\label{sec:Algorithm}
In this section, using the results provided in Section~\ref{sec:state_estimate}, we introduce the EXPress line failure detection using partially ObSErved information (EXPOSE) Algorithm. The EXPOSE Algorithm is summarized in Algorithm~\ref{algorithm:EXPOSE}.

Notice that for solving (\ref{eq:v_recovery_matrix}), (\ref{eq:LP_fail_detect}), and (\ref{eq:conv_Simul_v}) only the voltages of the nodes that are at most one hop away from the nodes in $\HH$ are required. 
Hence, not only the running time of the EXPOSE algorithm is independent of the number of line failures, it is independent of the size of the entire grid as well. This makes the EXPOSE Algorithm suitable for detecting line failures in large networks.


\begin{algorithm}[t]
\caption{Brute Force Search (BFS)}
\label{algorithm:BFS}
\small
\begin{trivlist}
\item\textbf{Input:} A connected graph $\GG$, attacked zone $\HH$, $V$, $V'_{\bar{\HH}}$, and $S'$ 
\end{trivlist}
\begin{algorithmic}[1]
\FOR {Any $\FF^{\dagger}\subseteq \LL_{\HH}$}
\STATE Compute $V^{\dagger}$ after removing the lines in $\FF^{\dagger}$ from $\GG$
\STATE Compute $e_{\FF^{\dagger}}\!=\!\|\re(V_{\bar{\HH}}^{\dagger}-V_{\bar{\HH}}')\|_2\!+\! \|\im(V_{\bar{\HH}}^{\dagger}-V_{\bar{\HH}}')\|_2$
\ENDFOR
\STATE Select $\FF = \argmin_{\FF^{\dagger}\subseteq \LL_{\HH}} e_{\FF^{\dagger}}$
\RETURN $\FF$
\end{algorithmic}
\end{algorithm}
\section{Brute Force Algorithm}\label{sec:BruteForce}
In order to compare the performance of the EXPOSE Algorithm with the previous works that were mostly under the DC power flows, we introduce the Brute Force Search (BFS) Algorithm for detecting line failures after the attack. The BFS Algorithm considers all possible line failure scenarios and returns the most likelihood solution. This method is the naive version of the approaches used in~\cite{tate2008line,tate2009double,zhu2012sparse,zhao2012pmu,zhu2014phasor} in similar settings to detect line failures given a partial phase angle measurements under the DC power flow model.

The idea is to compute the voltages $V_{\bar{\HH}}^{\dagger}$ for any possible set of line failures $\FF^{\dagger}\subseteq \LL_{\HH}$ and detect the one that is closest to $V_{\bar{\HH}}'$  as the most likely failure as follows:
\begin{align}\label{eq:brute_force}
\FF = \argmin_{\FF^{\dagger}\subseteq \LL_{\HH}} \|\re(V_{\bar{\HH}}^{\dagger}-V_{\bar{\HH}}')\|_2+ \|\im(V_{\bar{\HH}}^{\dagger}-V_{\bar{\HH}}')\|_2.
\end{align}

The BFS Algrithm is summarized in Algorithm~\ref{algorithm:BFS}. Notice that the BFS Algorithm is exponentially slower than the EXPOSE Algorithm, since it requires to solve the AC power flow solutions $2^{|\LL_{\HH}|}$ times. Moreover, since it requires to solve the power flow equations for the entire grid, in oppose to the EXPOSE Algorithm, its running time increases polynomially with the size of the grid.

The main shortcoming of the BFS Algorithm is its intractability for large networks. One way to speed up the BFS Algorithm is to stop whenever the $e_{\FF^{\dagger}}$ (as defined in line 3 of the algorithm) is less than a threshold. This may speed up the process but does not solve the intractability issue.

\section{Numerical Results}\label{sec:NumRes}
\begin{figure}[t]
\vspace*{-0.4cm}
\centering
\includegraphics[scale=0.3]{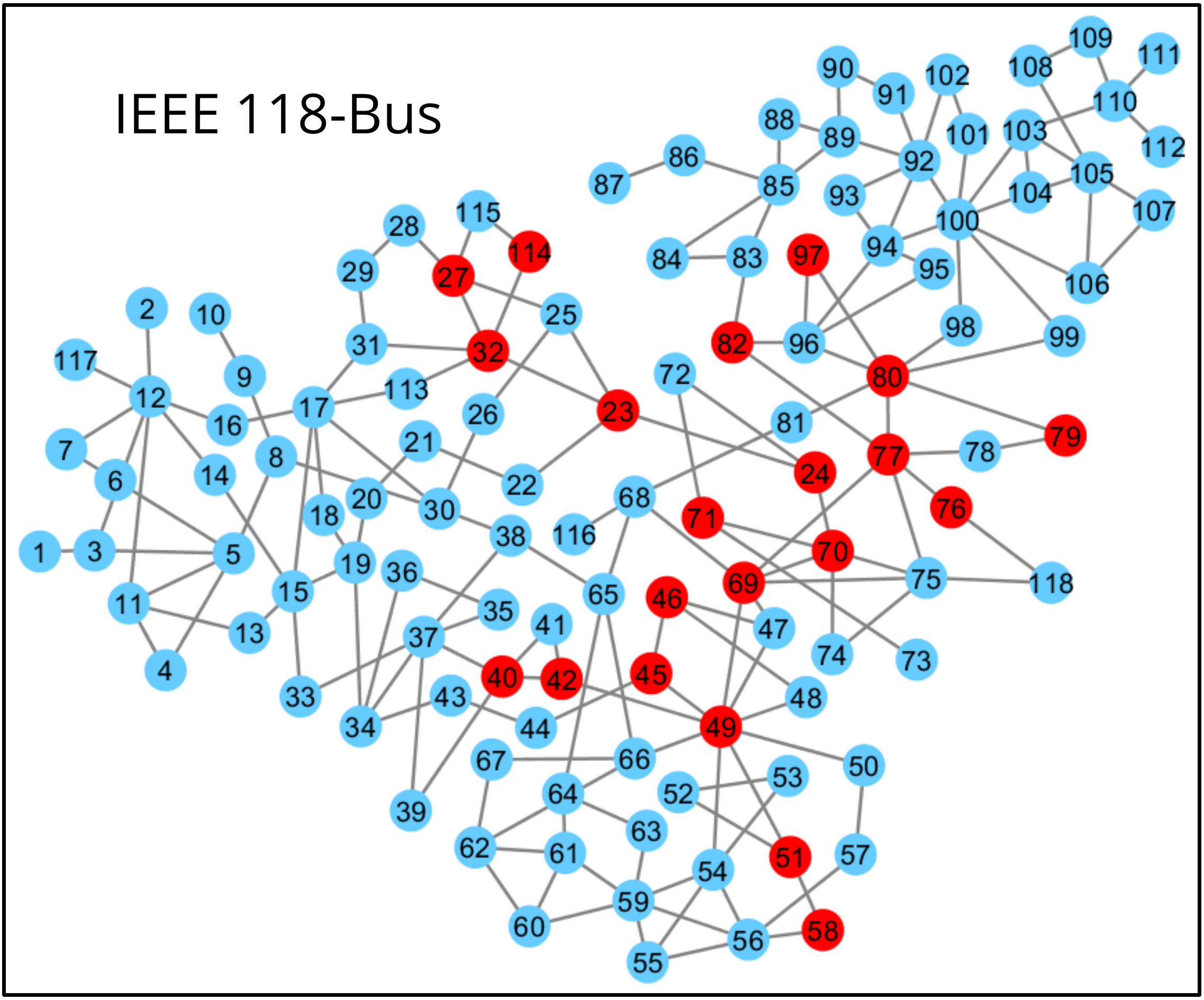}
\caption{An attacked zone used in~\cite{soltan2017power}, shown by red nodes, in the IEEE 118-bus system.}
\label{fig:attack_118MA}
\end{figure}

\begin{figure}[t]
\vspace*{-0.5cm}
\centering
\begin{subfigure}{0.24\textwidth}
\centering
\includegraphics[scale=0.32]{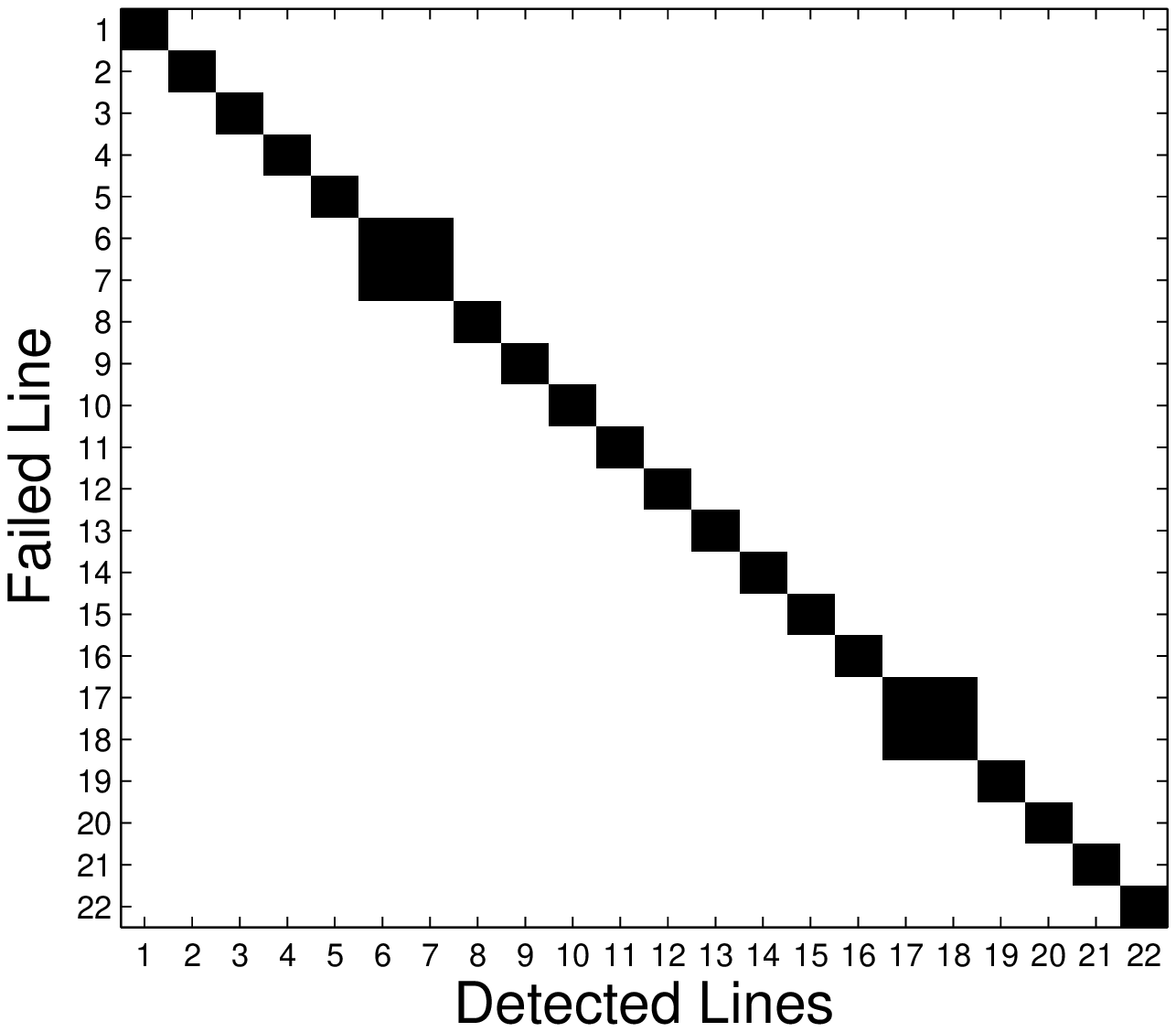}
\vspace*{-0.6cm}
\caption{}
\label{fig:1fail_118}
\end{subfigure}
\begin{subfigure}{0.24\textwidth}
\centering
\includegraphics[scale=0.32]{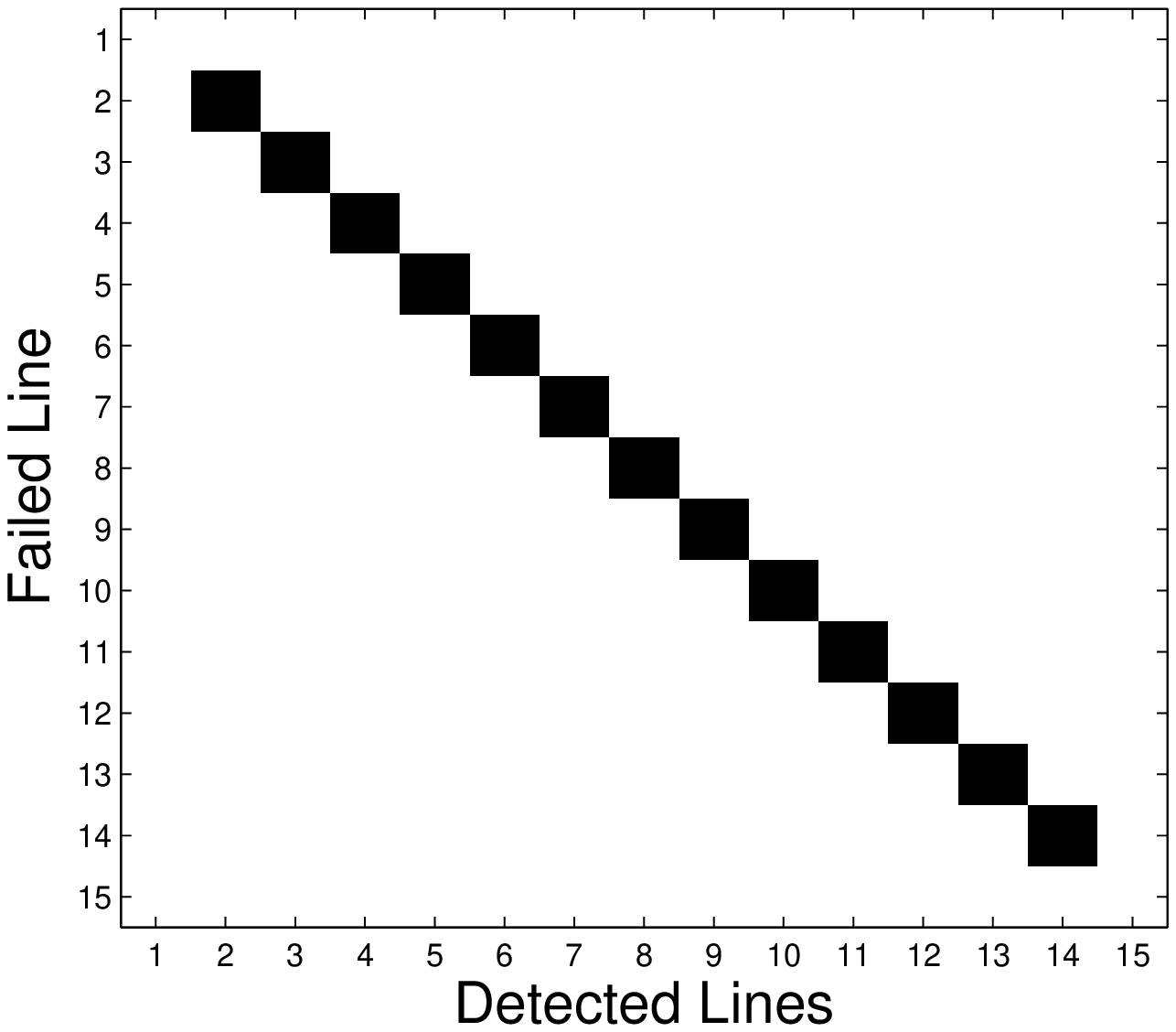}
\vspace*{-0.6cm}
\caption{}
\label{fig:1fail_300}
\end{subfigure}
\vspace*{-0.1cm}
\caption{Detected line failures after all possible single line failures using the EXPOSE Algorithm in (a) the zone shown in Fig.~\ref{fig:attack_118MA} in the IEEE 118-bus system, and (b) the zone shown in Fig.~\ref{fig:attack_300MA} in the IEEE 300-bus system.}
\label{fig:heatmap}
\end{figure}

\begin{figure}[t]
\vspace*{-0.5cm}
\centering
\begin{subfigure}{0.24\textwidth}
\includegraphics[scale=0.3]{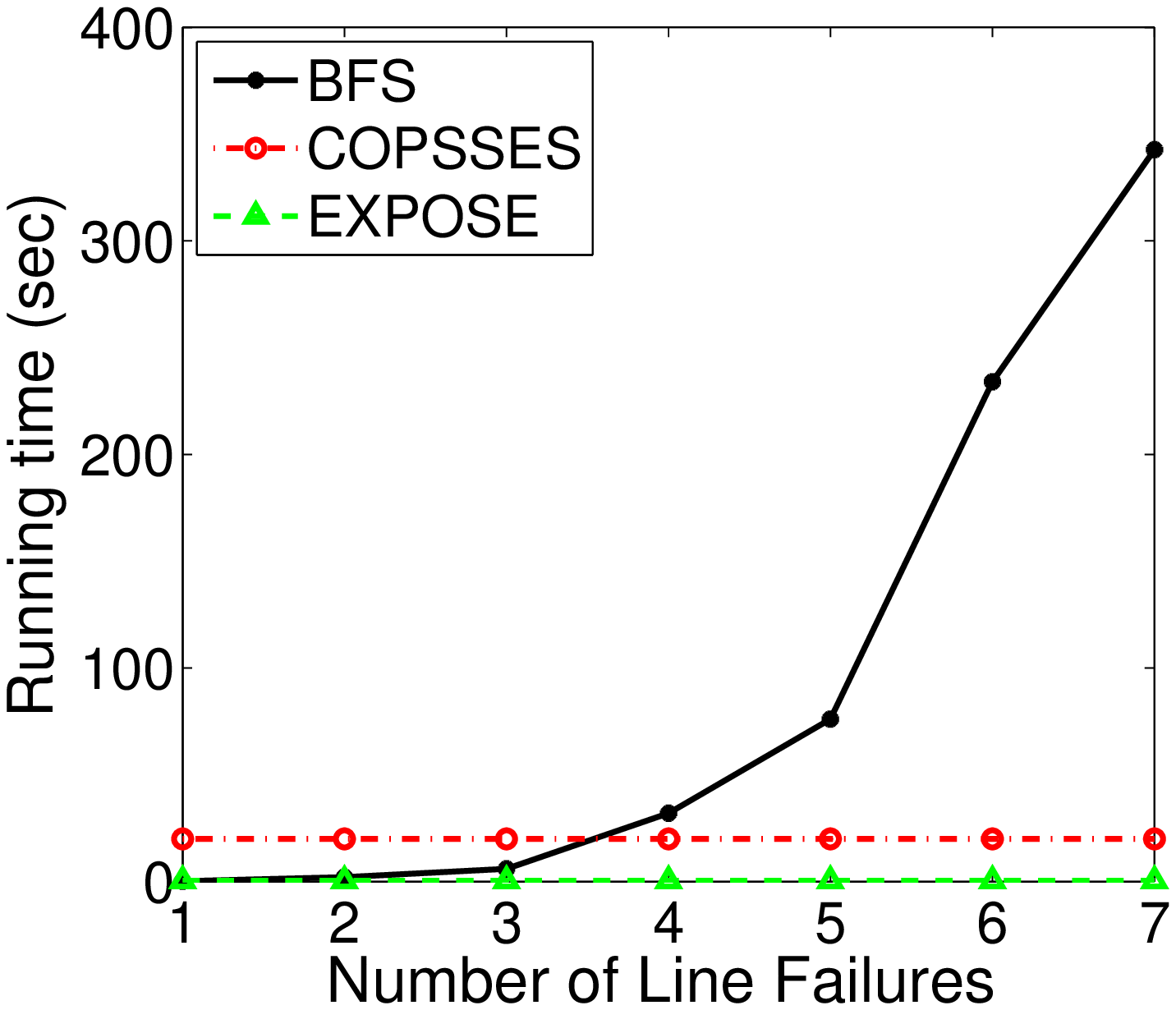}
\vspace*{-0.6cm}
\caption{}
\label{fig:RunningTime_M_A}
\end{subfigure}
\begin{subfigure}{0.24\textwidth}
\includegraphics[scale=0.3]{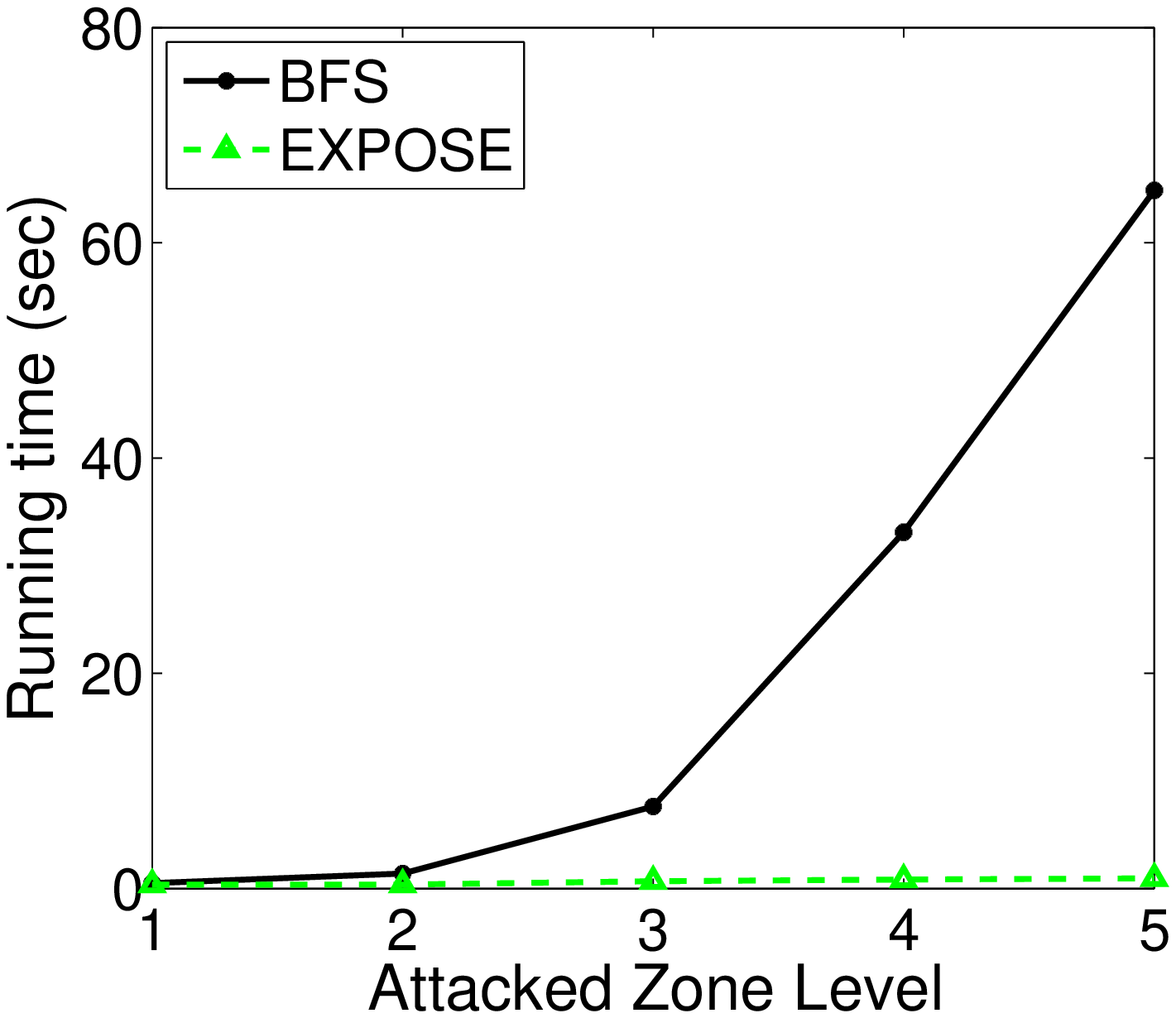}
\vspace*{-0.6cm}
\caption{}
\label{fig:RunningTime_Attacked_Area}
\end{subfigure}
\vspace*{-0.1cm}
\caption{Average running times of different line failures detection algorithms versus (a) the total number of line failures in the attacked zone shown in Fig.~\ref{fig:attack_300MA}, and (b) the size of the attacked zone as shown in Fig.~\ref{fig:attack_300gradual} in detecting triple line failures.}
\end{figure}

While in Section~\ref{sec:state_estimate}, we analytically proved that the EXPOSE Algorithm guarantees to accurately recover the voltages and detect line failures under some conditions (i.e., matched and acyclic attacked zones), in this section, we numerically confirm those results and also evaluate the EXPOSE Algorithm's performance when those conditions do not hold (i.e., general attacked zones).
\vspace*{-0.4cm}
\subsection{Matched and acyclic attacked zones}\label{subsec:MASim}

As we mentioned in Section~\ref{sec:Related}, to the best of our knowledge, our work in~\cite{soltan2017power} is the only other method for information recovery under the AC power flows that can be used to detect any number of line failures and scales well with the size of the grid. In~\cite{soltan2017power}, we introduced the Convex OPtimization for Statistical State
EStimation (COPSSES) Algorithm  and demonstrated that when the attacked zone is matched and acyclic  (i.e., matrices $\MM$ and $\D_{\HH}$ have full column rank), it can detect line failures with few errors. The COPSSES Algorithm uses a relaxation of the methods introduced in~\cite{soltan2016state}, which were based on DC power flow equations, for information recovery under the AC power flow equations. The advantage of the COPSSES Algorithm is that similar to the EXPOSE Algorithm, its running time is independent of the number of line failures.

\emph{In order to demonstrate the superiority of the EXPOSE Algorithm in this case, here we compare its performance and running time to the COPSSES Algorithm in addition to the BFS Algorithm.} For comparison purposes, we consider attacks on the same zones as considered in~\cite{soltan2017power} within the IEEE 118- and 300-bus systems. The zones are depicted in Figs.~\ref{fig:attack_118MA} and \ref{fig:attack_300MA}.

Recall from subsections~\ref{subsec:voltage_recover} and \ref{subsec:detect_lines} that when matrices $\MM$ and $\D_{\HH}$ have full column rank, as it is the cases here, the EXPOSE Algorithm can recover the voltages and detect the line failures accurately. Hence as we expected and can be seen in Fig.~\ref{fig:heatmap}, all the single line failures can be exactly detected using the EXPOSE Algorithm in the selected attacked zones within the IEEE 118- and 300-bus systems. Notice that the false positives in failures of lines 6 and 7 as well as 17 and 18 in the IEEE 118-bus system are due to the violation of the acyclicity of the attacked zone. Lines 6 and 7 (and also 17 and 18) are parallel lines that form a cycle with two nodes.

Moreover, lack of any detections after failures in lines 1 and 15 within that attacked zone in the 300-bus system is due to the fact that the AC power flows did not have a solution after those failures. Therefore those cases did not considered in evaluation of the EXPOSE Algorithm.

We considered up to 7 line failures in the zone depicted in Fig.~\ref{fig:attack_300MA}. In all the cases, as we expected, the EXPOSE Algorithm could exactly detect the line failures. The BFS Algorithm could also detect the line failures exactly in those scenarios. However, as it was shown in~\cite{soltan2017power}, the COPSSES Algorithm may result in few false positives and negatives in detecting single line failures, and more false positives and negatives as the number of line failures increases.

Fig.~\ref{fig:RunningTime_M_A} compares the running times of the three Algorithms in detecting line failures versus the number of line failures. As can be seen since the running times of the EXPOSE and COPSSES Algorithms are independent of the number of line failures, they both provide a constant running time as the number of line failures increases. However, as can be seen, the running time of the BFS Algorithm increases exponentially as the total number of line failures increases.

Overall, when the attacked zone is matched and acyclic, the EXPOSE Algorithm detects line failures as accurately as the BFS Algorithm, but exponentially faster.



\begin{figure}[t]
\centering
\begin{subfigure}{0.24\textwidth}
\includegraphics[scale=0.32]{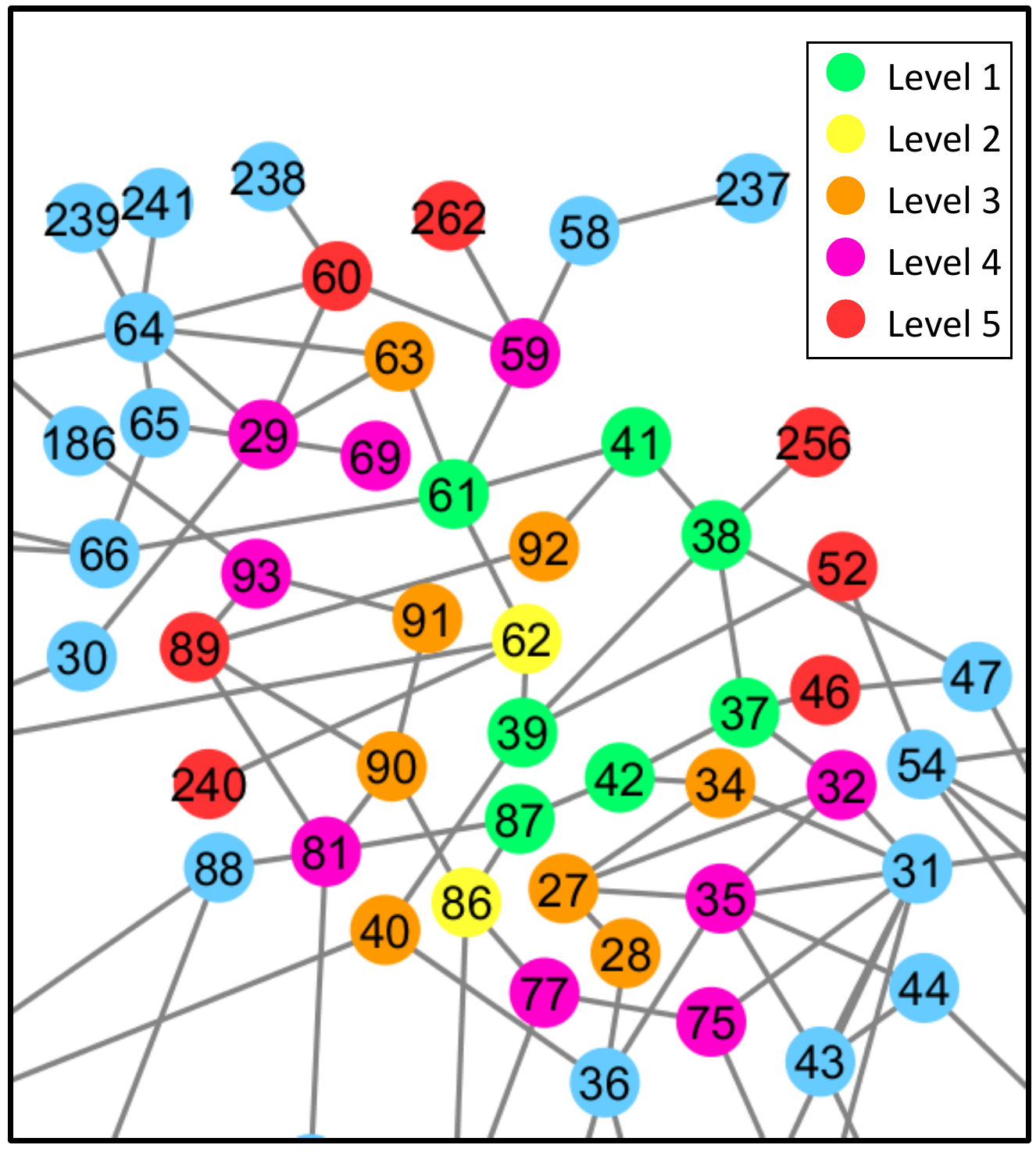}
\caption{}
\label{fig:attack_300gradual}
\end{subfigure}
\begin{subfigure}{0.24\textwidth}
\includegraphics[scale=0.3]{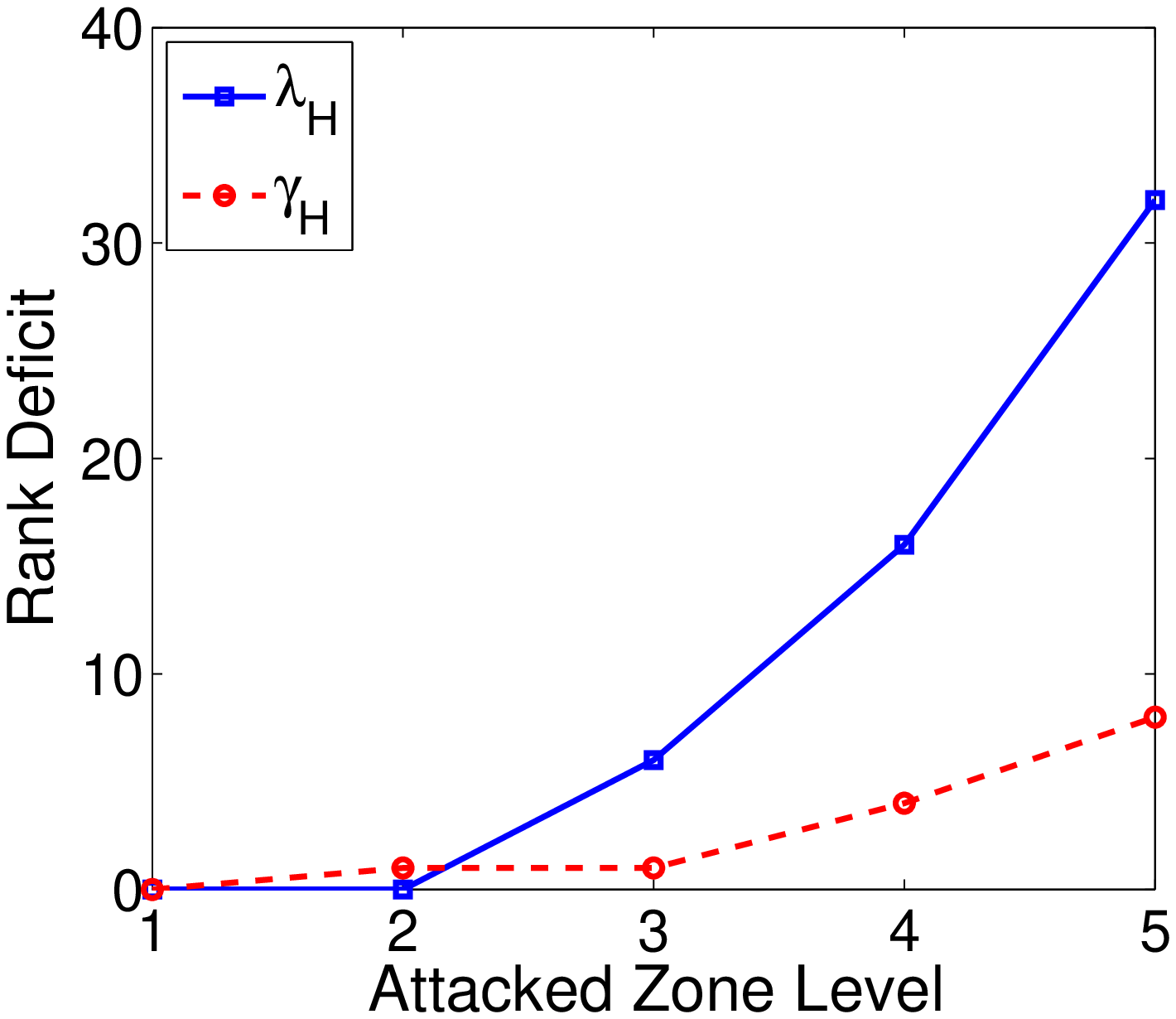}
\caption{}
\label{fig:RankDeficit}
\end{subfigure}
\caption{Nested attacked zones in the IEEE 300-bus system. (a) Nodes corresponding to different levels are shown in different colors, and (b) the rank deficit in the attacked zones in different levels.}
\end{figure}

\vspace*{-0.4cm}
\subsection{General attacked zones}\label{subsec:GenSim}

In order to evaluate the performance of the EXPOSE Algorithm as the attacked zone becomes larger and topologically more complex, in this subsection, we consider 5 nested attacked zones as depicted in Fig.~\ref{fig:attack_300gradual}. We denote the nodes that are added to the attacked zone at $i^{th}$ step by the \emph{level $i$} nodes. The level $i$ attacked zone is an attacked zone that consists of all the nodes in levels 1 to $i$.

As we proved in Section~\ref{sec:state_estimate} and briefly showed in Subsection~\ref{subsec:MASim}, when matrices $\MM$ and $\D_{\HH}$ have full column rank, then the EXPOSE Algorithm can recover the voltages and detects the line failures accurately. In order to show how far or close the topological properties of an attacked zone are to these conditions, we define $\lambda_{\HH}$ and $\gamma_{\HH}$ as follows:
\begin{align*}
\lambda_{\HH} &:= 2n_{\HH}-\rank(\MM),\\
\gamma_{\HH} &:= m_{\HH}-\rank(\D_{\HH}).
\end{align*}
It can be verified that when matrices $\MM$ and $\D_{\HH}$ have full column rank, then $\lambda_{\HH}=0$ and $\gamma_{\HH}=0$, respectively. Hence, $\lambda_{\HH}$ and $\gamma_{\HH}$ indicate the \emph{rank deficit} of matrices $\MM$ and $\D_{\HH}$.

Fig.~\ref{fig:RankDeficit} shows the $\lambda_{\HH}$ and $\gamma_{\HH}$ values for the different attacked zone levels. As can be seen, both values grow significantly in level 4 and 5 attacked zones. This means that the data outside of the attacked area is very insufficient to accurately detect the line failures based on the EXPOSE Algorithm in those levels.

First, in order to show the advantage of the EXPOSE Algorithm over brute force type algorithms, in Fig.~\ref{fig:RunningTime_Attacked_Area}, we compare the increase in the running times of the BFS and the EXPOSE Algorithms in detecting triple line failures as the number of nodes and lines increases in different levels. As can be seen in Fig.~\ref{fig:RunningTime_Attacked_Area}, the running time of the BSF Algorithm exponentially increases with the size of the attacked zone whereas that of the EXPOSE Algorithm only slightly increases. This along with Fig.~\ref{fig:RunningTime_M_A} clearly indicates that the BFS Algorithm (and algorithms with similar approaches) do not scale well with the size of the attacked zone and the number of line failures.

\begin{figure}[t]
\vspace*{-0.2cm}
\centering
\begin{subfigure}{0.24\textwidth}
\centering
\includegraphics[scale=0.3]{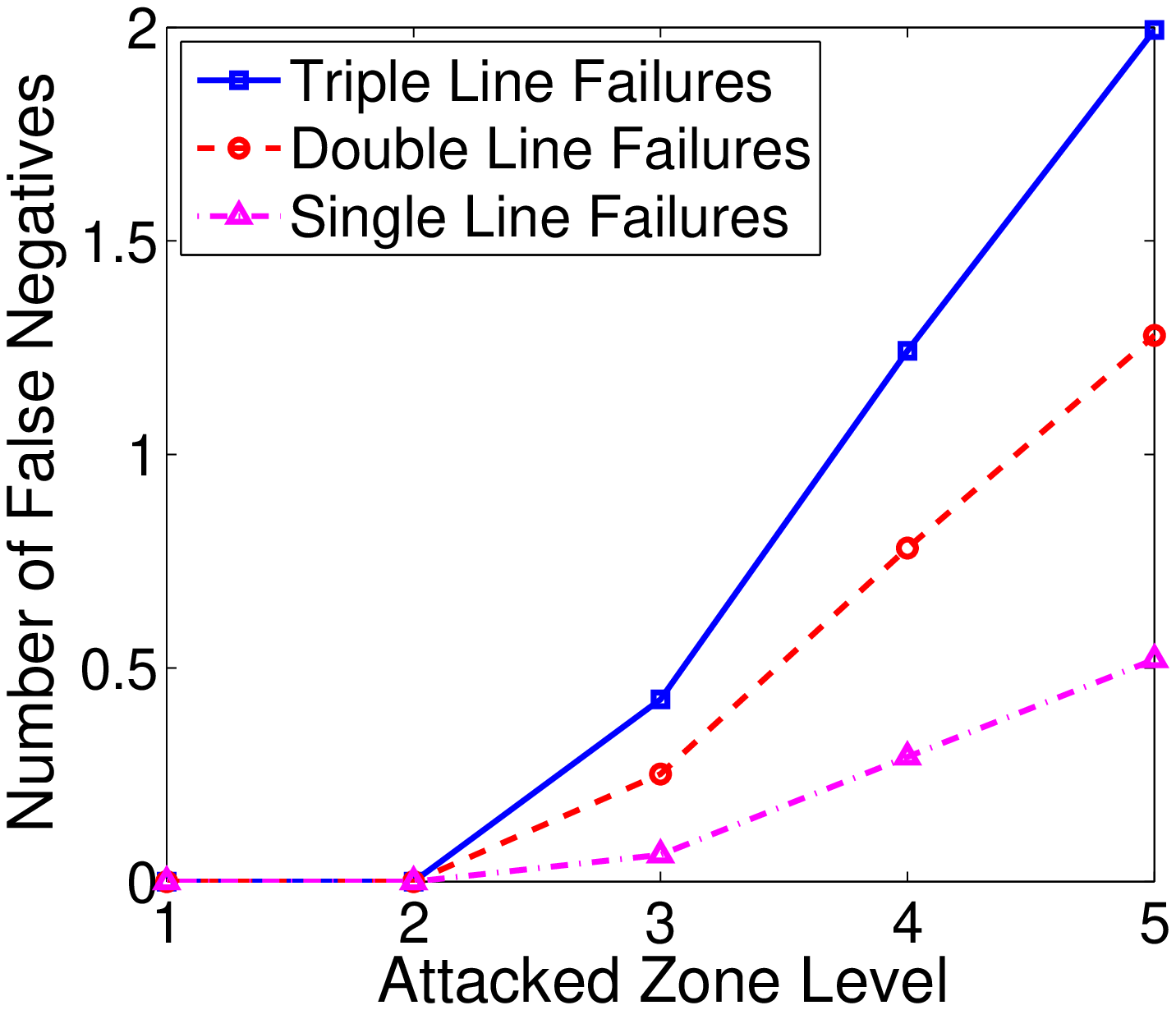}
\vspace*{-0.5cm}
\caption{}
\label{fig:FN}
\end{subfigure}
\begin{subfigure}{0.24\textwidth}
\centering
\includegraphics[scale=0.3]{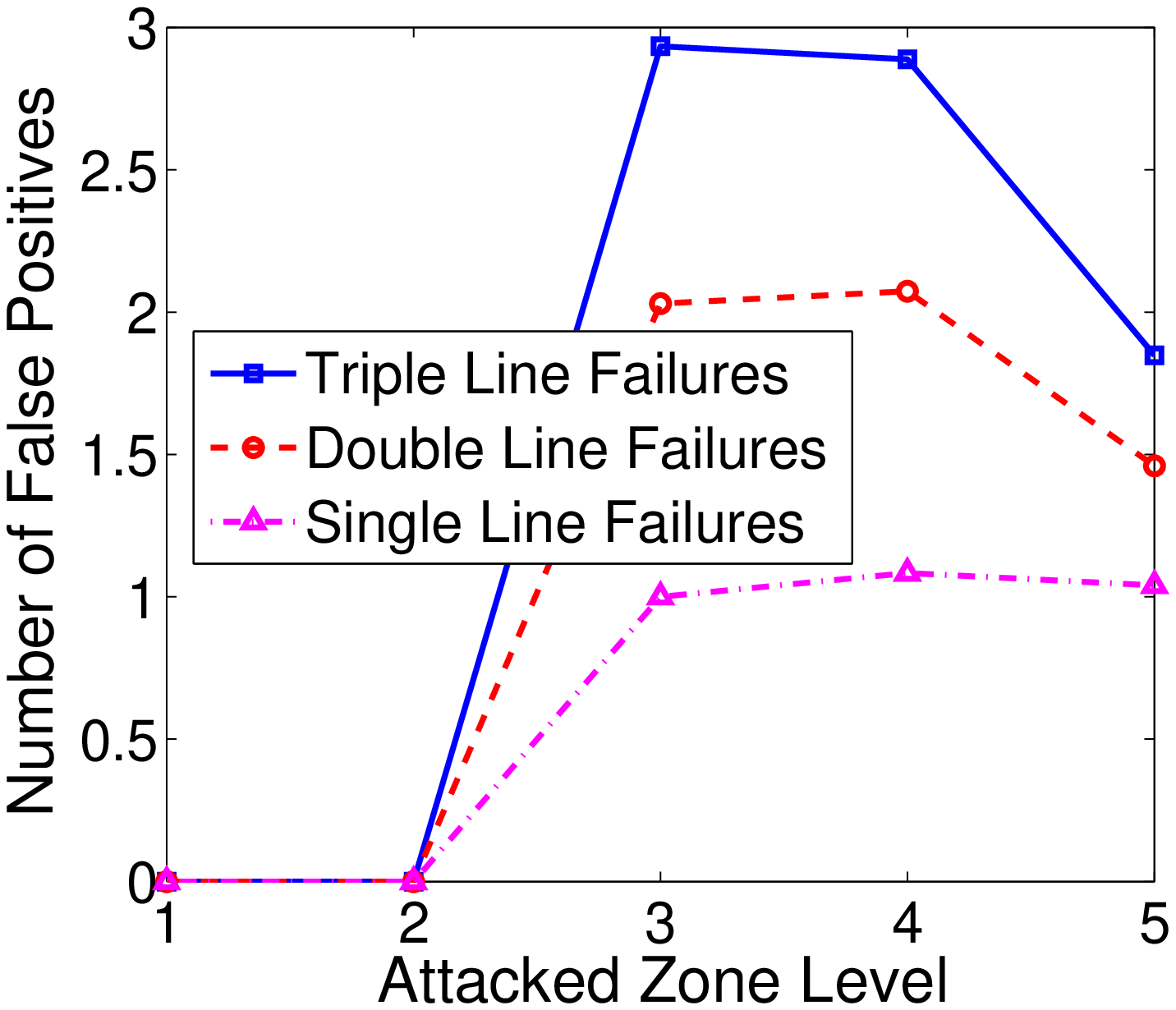}
\vspace*{-0.5cm}
\caption{}
\label{fig:FP}
\end{subfigure}
\begin{subfigure}{0.24\textwidth}
\vspace*{-0.07cm}
\centering
\includegraphics[scale=0.3]{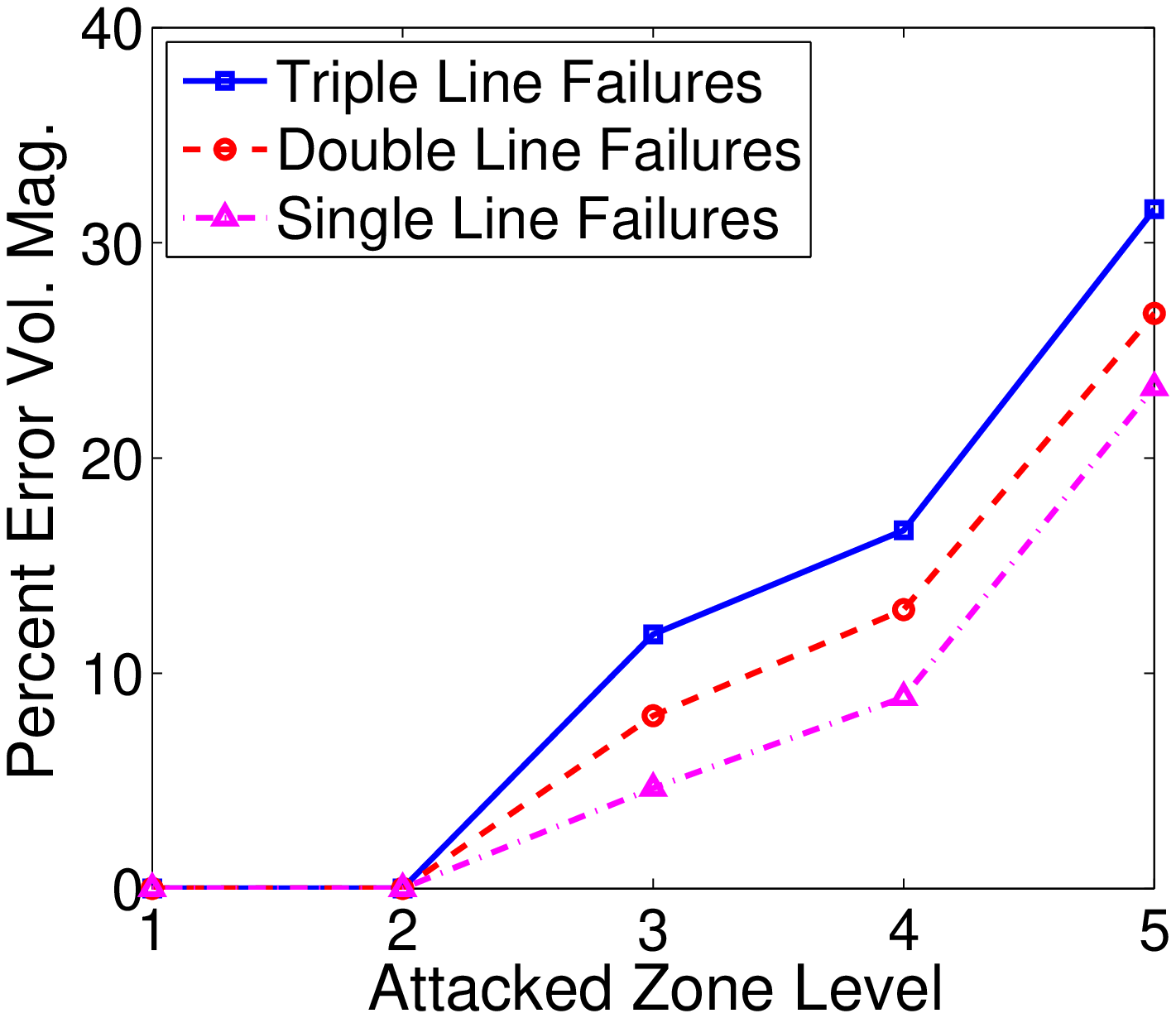}
\vspace*{-0.5cm}
\caption{}
\label{fig:VoltageError}
\end{subfigure}
\begin{subfigure}{0.24\textwidth}
\vspace*{-0.07cm}
\centering
\includegraphics[scale=0.3]{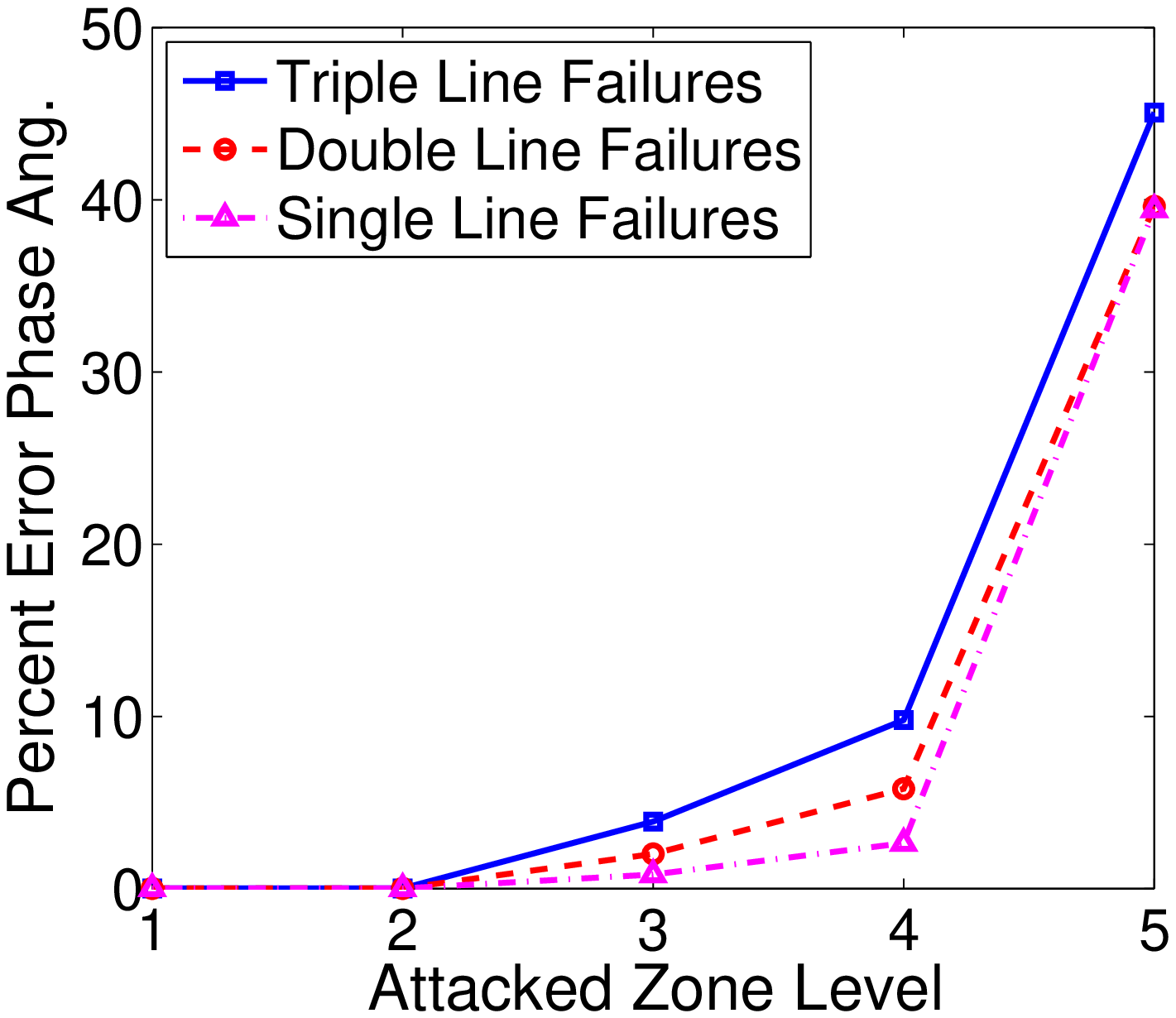}
\vspace*{-0.5cm}
\caption{}
\label{fig:VoltagePhase}
\end{subfigure}
\begin{subfigure}{0.24\textwidth}
\vspace*{-0.07cm}
\centering
\includegraphics[scale=0.3]{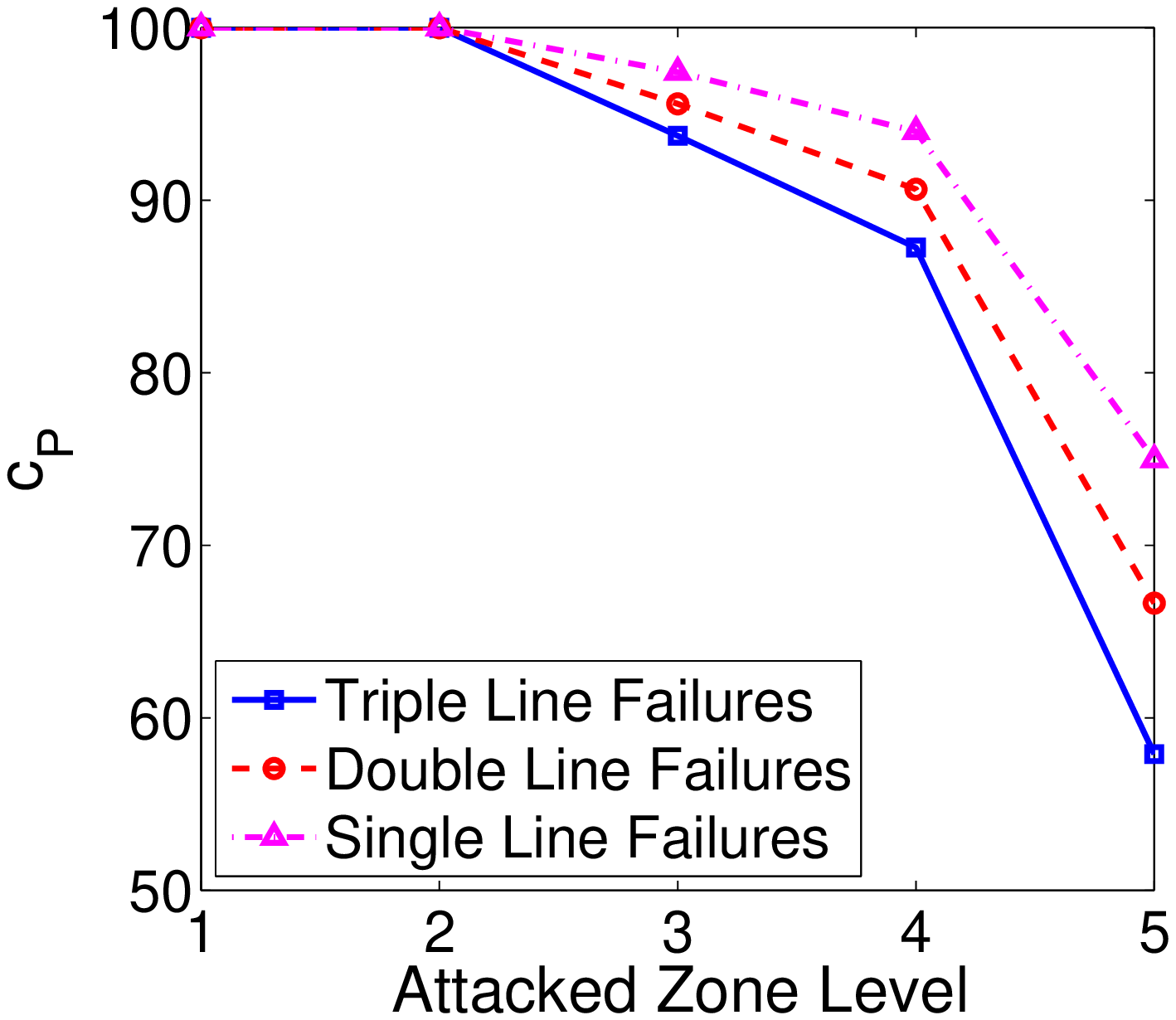}
\vspace*{-0.5cm}
\caption{}
\label{fig:TestP}
\end{subfigure}
\begin{subfigure}{0.24\textwidth}
\vspace*{-0.07cm}
\centering
\includegraphics[scale=0.3]{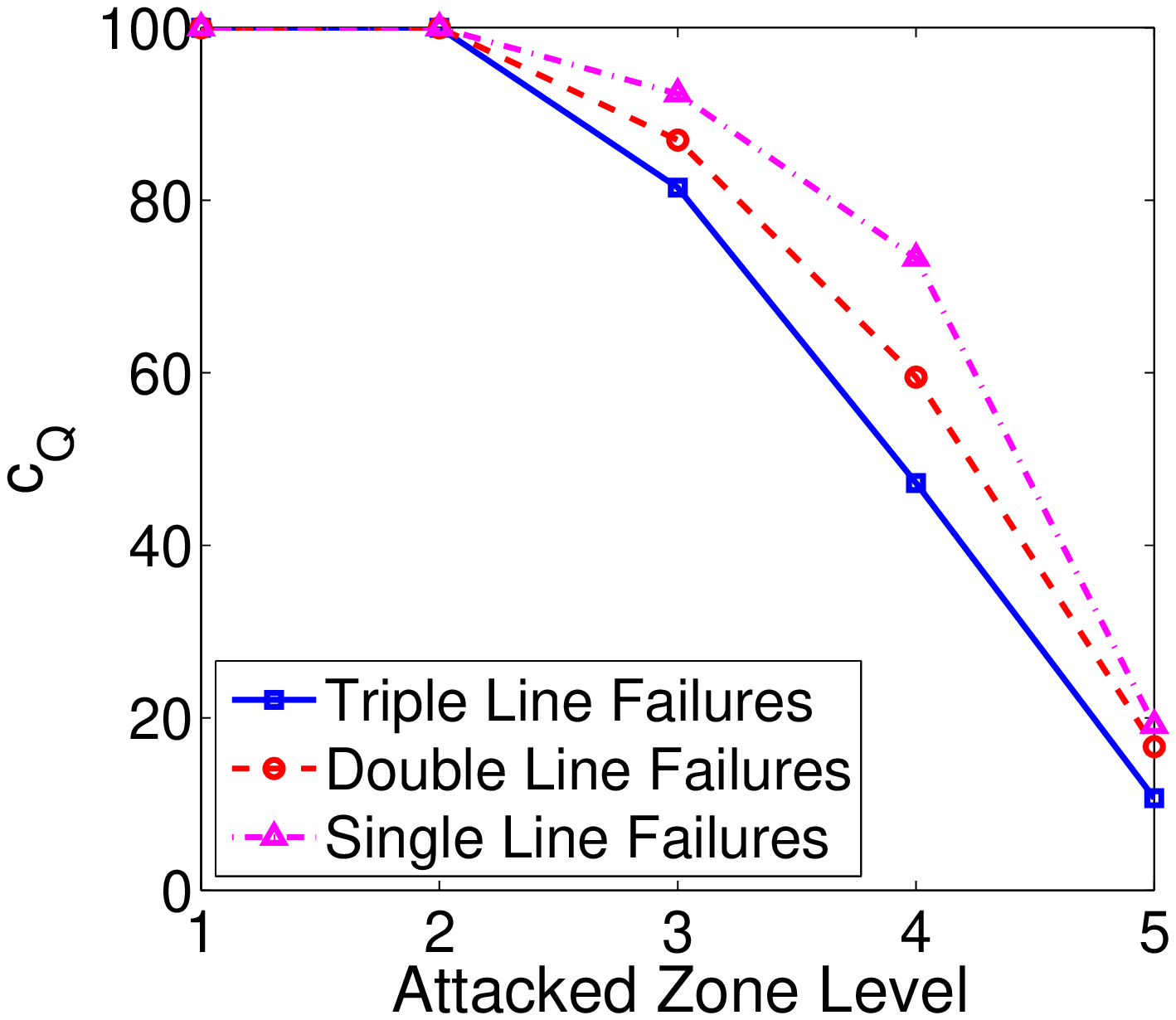}
\vspace*{-0.5cm}
\caption{}
\label{fig:TestQ}
\end{subfigure}
\caption{The EXPOSE Algorithm's performance after all single, double, and triple line failures versus the size of the attacked zone as shown in Fig.~\ref{fig:attack_300gradual}. (a) Average number of false negatives, (b) average number of false positives, (c) average percentage error in recovered voltage magnitudes, (d) average percentage error in recovered voltage phase angles, (e) average confidence of the solutions ($c_P$), and (f) average confidence of the solutions ($c_Q$).}
\label{fig:evaluation}
\end{figure}

%
%

In order to evaluate the performance of the EXPOSE Algorithm, we consider all single, double, and triple line failures in the nested zones. The results are presented in Fig.~\ref{fig:evaluation}.

The average number of false negatives and positives in detecting line failures in different attacked zone levels for all single, double, and triple line failures are presented in Figs.~\ref{fig:FN} and \ref{fig:FP}. It can be seen that as we expected for the level 1 attacked zone, there are no false negatives or positive. For the level 2 attacked zone also, although $\D_{\HH}$ does not have full column rank (see Fig.~\ref{fig:RankDeficit}), the EXPOSE Algorithm can still detect the line failures accurately. However, as the attacked zone becomes larger in higher levels, and $\lambda_{\HH}$ and $\gamma_{\HH}$ increase, we observe that the EXPOSE Algorithm results in false positives and negatives. An important observation here is that the EXPOSE Algorithm results on average in more false positives than negatives. This is a good characteristic of the EXPOSE Algorithm, since it means that by having an extra brute force search step on the detected line failures set, one can reduce number of false positives significantly.

The average error in recovered voltages in different attacked zone levels using the EXPOSE Algorithm for all single, double, and triple line failures are presented in Figs.~\ref{fig:VoltageError} and \ref{fig:VoltagePhase}. As can be seen, similar to the line failures detection, the EXPOSE Algorithm recovers the voltages accurately for the level 1 and 2 attacked zones. Moreover, for the level 3 and 4 attacked zones, the EXPOSE Algorithm recovers the voltage magnitudes and phase angles with less than $15\%$ and $10\%$ error, respectively. However, for the level 5 attacked zone, since $\lambda_{\HH}$ is too high (see Fig.~\ref{fig:RankDeficit}), the EXPOSE Algorithm  results in around $30\%$ and $40\%$ error in the recovered voltage magnitudes and phase angles, respectively.

Finally, as we introduced $c_P$ and $c_Q$ in subsection~\ref{subsec:conf_sol}, these metrics can be used to determine the confidence of the solutions obtained by the EXPOSE Algorithm. As can be seen in Figs.~\ref{fig:TestP} and \ref{fig:TestQ}, the $c_P$ and $c_Q$ values are directly correlated with the errors in voltages and number of false negatives. Hence, these values can effectively be used to compute the confidence of the solution obtained by the EXPOSE Algorithm. Notice that $c_Q$ is more sensitive than the $c_P$. Therefore, $c_Q$ can be used as the upper bound for the error and $c_P$ can be used as the lower bound.

We did not evaluate the performance of the BFS Algorithm here due to its very high running time (see Fig.~\ref{fig:RunningTime_Attacked_Area}). However, we expect that the BFS Algorithm could detect the line failures and recover the voltages with almost no error. Despite its accuracy, the BFS Algorithm is not practical for line failures detection in large networks. As we showed in this section, the EXPOSE Algorithm can provide relatively accurate results exponentially faster than the BFS Algorithm.
\section{Conclusion}\label{sec:Conclusion}
We studied cyber-physical attacks on power grids under the AC power flows. We leveraged the algebraic properties of the AC power flows to develop the EXPOSE Algorithm for detecting line failures and recovering the voltages after the attack.  We analytically proved that if the attacked zone has certain topological properties, the EXPOSE Algorithm can accurately recover the information. We also numerically demonstrated that in more complex attacked zones, it can still recover the information approximately well. The main advantages of the EXPOSE Algorithm are that its running time is independent of the size of the grid and number of line failures, and that it provides accurate information recovery under some conditions on the attacked zone. Moreover, it approximately recovers the information and provides the confidence of the solution when these conditions do not hold.

The results provided in this paper can be further used in different context as well. For example,  the EXPOSE Algorithm can be used to detect line failures when measurement devices are scarce and not ubiquitous. Moreover, the conditions on the attacked zone such that the EXPOSE Algorithm can accurately detect the line failures and recover the voltages, can be used for optimal measurement device placements in the grid. 

Despite it strengths, the EXPOSE Algorithm presented in this paper requires that the power system converges to a stable state after an attack. However, as the number of line failures increases, such an assumption may rarely holds. Therefore, the dynamics of the system after an attack should also be considered for an effective detection mechanism. Due to their complexity, study the dynamics of the power system after an attack is a very challenging task. Hence, exploring this and other directions is part of our future work. 
\section*{Acknowledgement}
This work was supported in part by DARPA RADICS under contract \#FA-8750-16-C-0054, funding from the U.S. DOE OE as part of the DOE Grid Modernization Initiative, and DTRA grant HDTRA1-13-1-0021.

\bibliographystyle{IEEEtran}
\bibliography{bib_AC}
\end{document}